\documentclass[%
 reprint,
 amsmath,
 amssymb,
 aps,
 prd,
 nofootinbib,
 superscriptaddress,
]{revtex4-2}
\usepackage{graphicx}
\usepackage{mathtools}
\usepackage{tabularray}
\UseTblrLibrary{booktabs}
\graphicspath{{./}{figures/}}
\usepackage{verbatim} 
\usepackage{bm}
\usepackage{hyperref}
\hypersetup{
    colorlinks=true,
    citecolor=blue,
    urlcolor=blue,
}
\newcommand{\rb}{\mathrm{b}}

\begin{document}

\title{Structure Formation and the Global 21-cm Signal in the Presence of Coulomb-like Dark Matter-Baryon Interactions}

\author{Trey Driskell}
\email{gdriskel@usc.edu}
\affiliation{Department of Physics $\&$ Astronomy, University of Southern California, Los Angeles, CA, 90007, USA}

\author{Ethan O.~Nadler}
\email{enadler@carnegiescience.edu}
\affiliation{Carnegie Observatories, 813 Santa Barbara Street, Pasadena, CA 91101, USA}
\affiliation{Department of Physics $\&$ Astronomy, University of Southern California, Los Angeles, CA, 90007, USA}

\author{Jordan Mirocha}
\affiliation{McGill Space Institute and McGill Physics Department, Montr\'eal, QC, H3A 2T8, Canada}

\author{Andrew Benson}
\affiliation{Carnegie Observatories, 813 Santa Barbara Street, Pasadena, CA 91101, USA}

\author{Kimberly K. Boddy}
\affiliation{Theory Group, Department of Physics, The University of Texas at Austin, Austin, TX 78712, USA}

\author{Timothy D. Morton}
\affiliation{Department of Physics $\&$ Astronomy, University of Southern California, Los Angeles, CA, 90007, USA}

\author{Jack Lashner}
\affiliation{Department of Physics $\&$ Astronomy, University of Southern California, Los Angeles, CA, 90007, USA}

\author{Rui An}
\affiliation{Department of Physics $\&$ Astronomy, University of Southern California, Los Angeles, CA, 90007, USA}

\author{Vera Gluscevic}
\email{vera.gluscevic@usc.edu}
\affiliation{Department of Physics $\&$ Astronomy, University of Southern California, Los Angeles, CA, 90007, USA}

\begin{abstract}
Many compelling dark matter (DM) scenarios feature Coulomb-like interactions between DM particles and baryons, in which the cross section for elastic scattering scales with relative particle velocity as $v^{-4}$. Previous studies have invoked such interactions to produce heat exchange between cold DM and baryons and alter the temperature evolution of hydrogen. In this study, we present a comprehensive study of the effects of Coulomb--like scattering on structure formation, in addition to the known effects on the thermal history of hydrogen. We find that interactions which significantly alter the temperature of hydrogen at Cosmic Dawn also dramatically suppress the formation of galaxies that source the Lyman-$\alpha$ background, further affecting the global 21-cm signal. In particular, an interaction cross section at the current observational upper limit leads to a decrease in the abundance of star-forming halos by a factor of $\sim 2$ at $z\sim 20$, relative to cold, collisionless DM. We also find that DM that is 100\% millicharged cannot reproduce the depth and the timing of the reported EDGES anomaly in any part of the parameter space. These results critically inform modeling of the global 21-cm signal and structure formation in cosmologies with DM--baryon scattering, with repercussions for future and upcoming cosmological data analysis.
\end{abstract}


\maketitle

\section{Introduction}\label{sec:intro}
Numerous cosmological and astrophysical observations testify to the existence of an additional matter-like component---dark matter (DM)---which outweighs the baryonic matter in the Universe~\cite{Bertone2004}. Thus far, the standard description of DM as a cold and collisionless (CDM) particle has been mostly successful at satisfying observational constraints, with some potential evidence that the CDM model could be incomplete \cite{Bullock2017, Tulin2017}. At the same time, many DM  models go beyond the CDM paradigm and invoke interactions with the Standard Model of particle physics, motivating the exploration of interacting DM (IDM) \cite{cosmicvisions2017,Boddy:2022knd}. 

Observations can probe IDM microphysics scenarios \cite{Gluscevic2019, peterbuckley2018} that are difficult to access with direct detection experiments \cite{Slatyer2009} and other laboratory searches, in particular light DM (with particle masses $m_{\chi}\lesssim 1\,  \mathrm{GeV}$) and DM that couples strongly with the Standard Model \cite{Lewin1995}. 
Additionally, cosmology is a versatile probe of DM physics, broadly sensitive to swaths of IDM models: regardless of the specific high-energy description of DM, elastic scattering between DM particles and baryons at low energies affects the thermal history of the Universe and the growth of matter perturbations on a range of physical scales. The relevant interaction physics is largely captured by parameterizing the momentum-transfer cross section as a power law of the relative particle velocity $v$ \cite{Sigurdson2004, Dvorkin2014, Boehm2004, Slatyer2018, Gluscevic2018,KBVG2018, Boddy2018, Xu:2021lmg, Nguyen2021, Maamari2020, Rogers2021, Becker2020,Nadler:2019zrb,DES:2020fxi, Li:2022mdj, Chen:2002yh, Boddy:2022tyt}. In this study, we focus on Coulomb-like interactions where $\sigma(v)=\sigma_{0}v^{-4}$, $\sigma_0$ is the unknown cross section prefactor, and DM scatters with all baryons. A subclass of these models includes millicharged DM which carries a fractional electrical charge and only scatters with charged particles. Such models can, for example, arise in scenarios featuring a dark photon mixing with the Standard Model photon \cite{Sigurdson2004, Barkana:2018qrx, McDermott:2010pa, Dvorkin:2019zdi, Berlin:2021kcm, Baryakhtar:2020rwy, Giovanetti:2021izc, Barkana:2018lgd}.

IDM models in which the interaction cross section decreases as the Universe expands and cools, and particle velocities redshift away, mainly suppress density perturbations, affecting the cosmic microwave background (CMB) anisotropy and leading to an under-abundance of small halos later on \cite[e.g.,][]{Nadler:2019zrb, Maamari2020}. Conversely, $v^{-4}$ scattering can continue to occur efficiently after recombination and alter conditions in the late-time intergalactic medium (IGM), affecting novel observables, such as the 21-cm signal from neutral hydrogen during Cosmic Dark Ages and Cosmic Dawn \cite{Tashiro2014, Munoz2015, Barkana:2018qrx, Slatyer2018, Fialkov2018, Short2022, Kovetz2018}. A number of experiments are currently working on measuring the sky-averaged 21-cm signal \cite{Singh2017, Bowman2018, Philip2019}, while others are pursuing power spectrum measurements or tomographic mapping at corresponding wavelengths \cite{Furlanetto:2004zw,Trott:2020szf,Mertens:2020llj,Ghosh:2010tz,HERA2021}. In fact, in 2018, the EDGES collaboration announced a measurement of an absorption feature in the sky-averaged radio spectrum, which would correspond to a 21-cm signal from atomic hydrogen at $z\sim 17$ \cite{Bowman2018}. Although the interpretation of this measurement is still controversial \cite{Barkana:2018qrx, Fialkov2018,Berlin2018, Munoz2018, Liu2019, Creque-Sarbinowski:2019mcm}, it is now clear that the global 21-cm signal can uniquely probe new aspects of DM microphysics \cite{Munoz:2019hjh}. 

As the first sources of light were turning on during Cosmic Dawn~\cite{FurlanettoPritchard2006}, the Lyman-$\alpha$ radiation background filled the IGM, adding to the radiative transitions in the atomic hydrogen populating the IGM. In particular, Lyman-$\alpha$ radiation mediated transitions between the two hyperfine ground states of neutral hydrogen, via the Wouthuysen-Field effect \cite{Wouthuysen1952,Field1958}, altering the 21-cm spin temperature, and driving it closer to the temperature of the gas. The competition between this effect and the heating from astrophysical sources and CMB photons is expected to produce the characteristic absorption feature against the CMB spectrum, at frequencies corresponding to the redshifted 21-cm signal from $z\sim 20$ \cite{Loeb:2006za}. 

The depth of the absorption trough is controlled by the Lyman-$\alpha$ budget and the temperature of the gas in the IGM. On the one hand, heat exchange between DM and hydrogen affects the temperature evolution of baryons \cite{Munoz2015,Barkana:2018lgd}. On the other, their momentum exchange affects the formation of structure, and specifically the formation of virialized DM halos \cite{Nadler:2019zrb}, which host the galaxies that source the Lyman-$\alpha$ background. While the effects of DM--baryon heat exchange on the 21-cm signal have been studied extensively, the effects of momentum exchange have not been studied in this context. In particular, all previous studies of the global 21-cm signal in the presence of DM--baryon interactions only account for the effects of IDM on the temperature evolution of baryons, and assume that the impact of interactions on structure formation is negligible. 

In this study, we demonstrate that accounting for the effects of both heat and momentum exchange is critical to modeling the 21-cm signal in the presence of DM--baryon scattering, where 100\% of DM interacts with baryons.
We also include the often-neglected scattering correction to the Wouthuysen–Field coupling strength from line profile effects, which is essential to model the coupling at the extremely low hydrogen temperatures ($\lesssim 1~\mathrm{K}$) achieved in IDM cosmologies. 

\begin{figure*}
    \centering
    \includegraphics[width=0.7\textwidth]{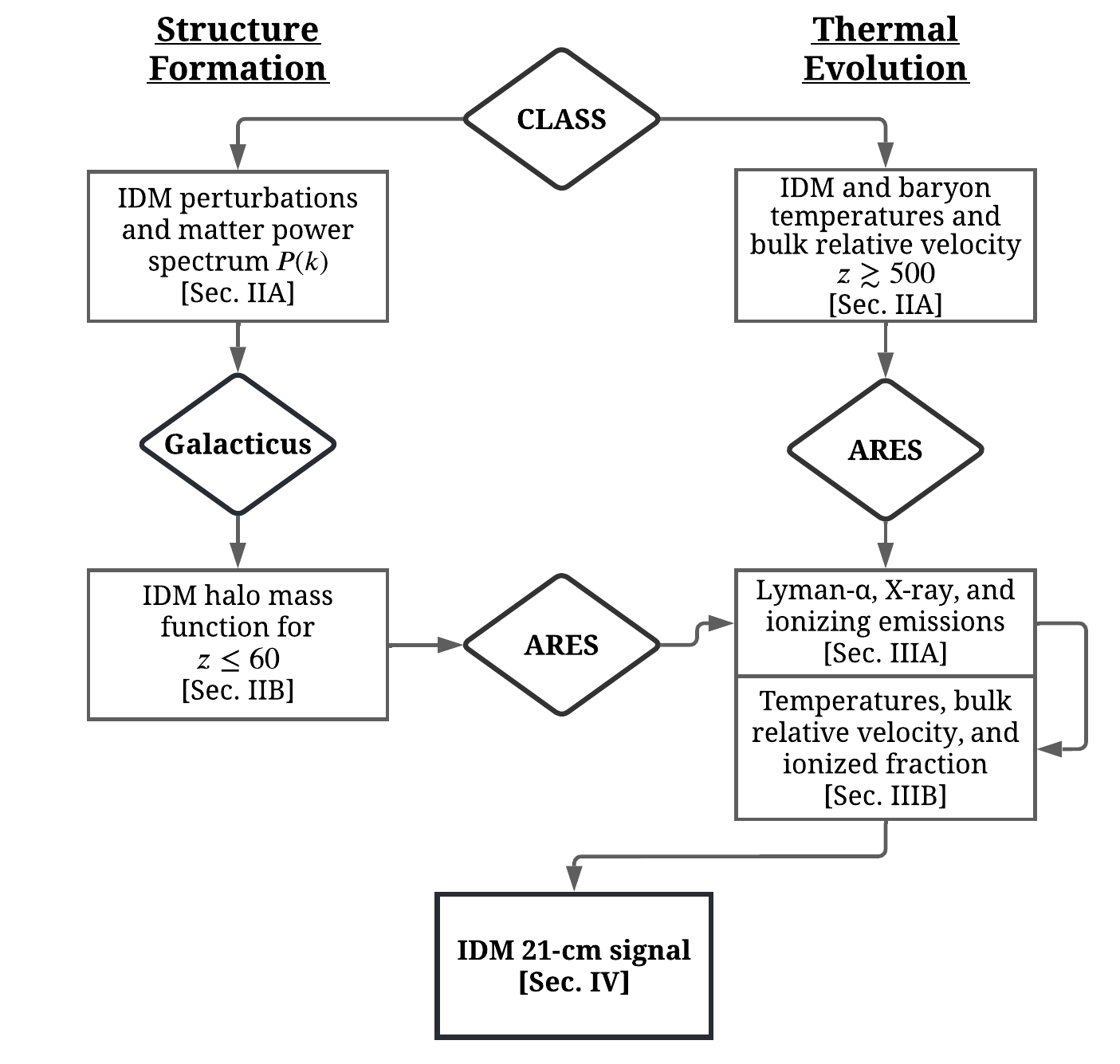}
    \caption{Our algorithm for modeling the global 21-cm signal in cosmologies with Coulomb-like DM-baryon scattering (IDM). We use a modified version of \texttt{CLASS} to compute the linear matter power spectrum $P(k)$ and evolve DM and baryon temperatures and bulk relative velocities at early times~(Sec.~\ref{subsec:linear}). The linear $P(k)$ serves as an input to \texttt{Galacticus}, which we use to compute the redshift-dependent halo mass function (Sec.~\ref{subsec:nonlinear}) in cosmologies with IDM. The halo mass function is then used in \texttt{ARES} to obtain the Lyman-$\alpha$ background, $X$--ray heating rate, and photo-ionization rate from astrophysical sources, consistent with IDM power suppression. The coupled post-recombination evolution of the bulk velocity and DM and baryon temperatures is then computed using a modified version of the \texttt{ARES} code. Finally, these quantities are used by \texttt{ARES} to predict the global 21-cm brightness temperature (Sec.~\ref{sec:21cm}). }
    \label{fig:pipeline}
\end{figure*}

In particular, we model the global 21-cm signal as a function of the DM particle mass $m_\chi$ and the interaction cross section prefactor $\sigma_0$. Our modeling approach is as follows. First, we use a modified version of the linear Boltzmann solver \texttt{CLASS}\footnote{\url{https://github.com/kboddy/class_public/tree/dmeff}} \cite{Blas:2011rf} to obtain the linear matter power spectrum, temperatures of DM and baryons, and their relative bulk velocity, consistent with a given choice of $m_\chi$ and $\sigma_0$. We use the output of \texttt{CLASS} at $z=500$ as an initial condition for the next stage of our calculations, stopping the linear solver at a redshift high enough to ensure that perturbations are still linear and that the thermal evolution has not yet been affected by emission from galaxies. We solve for subsequent thermal evolution using a modified version of the \texttt{ARES}\footnote{\url{https://github.com/mirochaj/ares}} code \cite{Mirocha2014,Mirocha2017}, which carefully models ionization history and heating emission background in the IGM; both ionization and heating affect IDM interactions, as they alter relative particle velocity and number density of targets for scattering. To model the formation of halos and galaxies consistent with IDM, we employ the open-source model \texttt{Galacticus}\footnote{\url{https://github.com/galacticusorg/galacticus}} \cite{Benson:2010kx} to predict the evolution of the halo mass function based on the extended Press-Sechter (ePS) formalism \cite{Press:1973iz,Bond:1990iw}. By combining an accurate model of thermal history and structure formation in interacting cosmologies, we employ \texttt{ARES} to perform a self-consistent calculation of the global 21-cm signal in IDM. Fig.~\ref{fig:pipeline} schematically illustrates our analysis algorithm.

We find that including the effects of the heat and momentum exchange and the corrected modeling of the Wouthuysen–Field effect significantly changes the global signal, compared to previous studies. In particular, for $v^{-4}$ interactions between DM and all baryons (i.e., including neutral targets), the general timing and depth of the EDGES signal is consistent with only a small window of parameter space with $10^{-42}\,\mathrm{cm^{2}} \lesssim \sigma_{0}\lesssim 10^{-41}\,\mathrm{cm^{2}}$, allowed by the CMB measurements. However, we find that 100\% millicharged DM \textit{cannot} produce the signal reported by EDGES for any DM cross section or particle mass---regardless of any other bounds on this model. This finding has deep implications for self-consistent modeling of the IDM effects on structure formation and thermal evolution, highly relevant for extracting novel DM physics from future 21-cm and large scale structure data.

This paper is structured as follows. Sec.~\ref{sec:mc_struct} describes the effects of Coulomb-like DM--baryon scattering on structure formation the baryonic temperature evolution. Sec.~\ref{sec:21cm} details our approach to modeling the global 21-cm signal in IDM cosmology. Sec.~\ref{sec:edges} presents our results, in particular identifying the regions of Coulomb-like and millicharged DM parameter space compatible with the EDGES measurement. We discuss caveats of our modeling and general implications of $v^{-4}$ DM--baryon interactions for structure formation in Sec.~\ref{sec:discussion}, and we summarize in Sec.~\ref{sec:conclusion}.

Throughout, we assume baseline cosmological parameters that match the best-fit values from \textit{Planck} 2018 \cite{Planck2018}: $h=0.6766$, $\Omega_{\mathrm{b}}h^{2}=0.0224$, $\Omega_{\chi}h^{2}=0.119$, $\mathrm{Log}(10^{10}A_{\mathrm{s}})=3.043$, $n_{\mathrm{s}}=0.965$, $Y_{\mathrm{He}}=0.245$, $\tau_{\mathrm{reio}}=0.0540$, and $N_{\mathrm{eff}}=3.046$.

\section{Structure Formation with Coulomb-like DM--Baryon Interactions}\label{sec:mc_struct}

We focus on Coulomb-like DM--baryon elastic scattering, parameterized by a momentum-transfer cross section, $\sigma(v)=\sigma_{0}v^{-4}$, where $v$ is the relative particle velocity and $\sigma_{0}$ is its amplitude, and the target mass is that of an average baryonic particle. We also consider millicharged DM, which only interacts with charged particles (in our case, free protons and electrons). In both cases, we assume that $100\%$ of the DM interacts with baryons.

In Sec.~\ref{subsec:linear}, we summarize the linear Boltzmann equations describing the evolution of density perturbations, which we solve to obtain the linear matter power spectrum $P(k)$. We also present the equations that capture the evolution of baryon and DM temperatures, and their relative bulk velocities. In Sec.~\ref{subsec:nonlinear}, we model the non-linear growth of perturbations in IDM cosmology and the resulting halo mass functions (HMFs). These HMFs feed into our calculation of the the sky-averaged 21-cm brightness temperature in Sec.~\ref{sec:21cm}.

\subsection{Linear Evolution}
\label{subsec:linear}
\subsubsection{Modified Boltzmann Equations}
\begin{figure}[t]
    \centering
    \includegraphics[width=\linewidth]{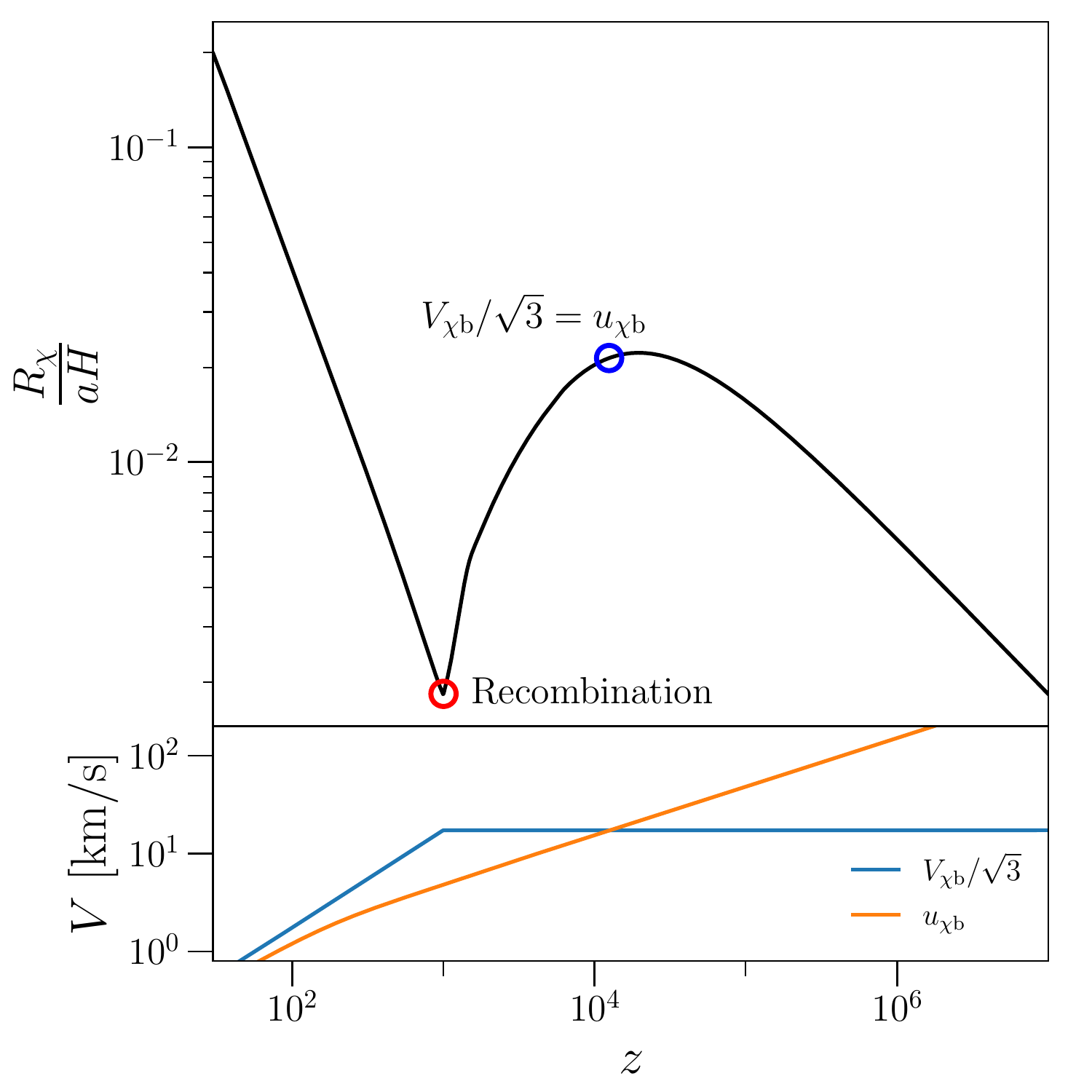}
    \caption{The momentum-transfer rate $R_\chi$ divided by the Hubble expansion rate $aH$, as a function of redshift, for a Coulomb-like DM--proton interaction model with $\sigma=\sigma_{0}v^{-4}$, where $\sigma_{0}=10^{-41}\ \mathrm{cm^{2}}$ (approximately equal to the current upper bound from \textit{Planck}) and DM particle mass $m_{\chi}=1\ \mathrm{MeV}$. Here, we set the scattering targets to be all baryons, and assume that all of DM is interacting. The red circle (left) denotes recombination, and the blue circle (right) marks the redshift when the bulk relative velocity exceeds the thermal velocity. The non-monotonic behavior of $R_\chi$ is driven by the fact that bulk relative velocity begins to dominate the relative particle velocity around the epoch of matter--radiation equality, and rapidly redshifts after recombination, as shown in the bottom panel.}
    \label{fig:Rchi}
\end{figure}

As shown in previous studies \cite[e.g.,][]{Sigurdson2004, Boehm2004, Dvorkin2014, Gluscevic2019, Boddy2018}, DM--baryon interactions introduce drag between the two fluids, coupling their temperatures and bulk relative velocities. The effects of interactions on perturbations is captured by a modification to the linear Boltzmann equations. Assuming the two fluids are non-relativistic, the Boltzmann equations in synchronous gauge read 
\begin{equation}\label{boltzmann}
    \begin{split}
        \dot{\delta}_{\chi} = &-\theta_{\chi} - \frac{\dot{h}}{2}, \qquad \dot{\delta}_{\rb} = -\theta_{\rb} - \frac{\dot{h}}{2}, \\ \dot{\theta}_{\chi} = &-\frac{\dot{a}}{a}\theta_{\chi}+c^{2}_{\chi}k^{2}\delta_{\chi} + R_{\chi}\left(\theta_{\rb}-\theta_{\chi}\right), \\
        \dot{\theta}_{\rb} = &-\frac{\dot{a}}{a}\theta_{\rb}+c^{2}_{\rb}k^{2}\delta_{\rb} + \frac{\rho_{\chi}}{\rho_{\mathrm{b}}}R_{\chi}\left(\theta_{\chi}-\theta_{\rb}\right) \\ 
        &+ R_{\gamma}\left(\theta_{\gamma}-\theta_{\rb}\right),
    \end{split}
\end{equation}
where $\delta_{\chi}$ and $\delta_{\rb}$ are the DM and baryon overdensities, $\theta_{\chi}$ and $\theta_{\rb}$ are the DM and baryon velocity divergences, $h$ is the trace of the scalar metric perturbation, $c_{\chi}$ and $c_{\rb}$ are the DM and baryon sound speeds, 
$R_{\gamma}$ is the rate coefficient of momentum transfer between baryons and photons due to Compton scattering, and $R_{\chi}$ is the rate coefficient of momentum transfer between DM and baryons, due to the interactions. The momentum-transfer rate coefficient for DM scattering off multiple baryon species is given by
\begin{equation}\label{Rchi}
    R_{\chi} = \sqrt{\frac{2}{\pi}} \frac{a}{3}\sum_{t}\frac{\rho_{\mathrm{t}}\sigma_{0}}{m_{\chi}+m_\mathrm{t}}\left(\frac{T_{\chi}}{m_{\chi}}+\frac{T_{\mathrm{K}}}{m_{\mathrm{t}}}+\frac{V^{2}_{\mathrm{rms}}}{3}\right)^{-\frac{3}{2}},
\end{equation}
where the sum is over the target baryon species, $m_{\mathrm{t}}$ is the mass of the target, and $\rho_{\mathrm{t}}$ is the density of the target, $T_{\chi}$ is the DM temperature, and $T_{\mathrm{K}}$ is the baryon temperature\footnote{The symbol $T_{\rb}$ is more commonly used for the baryon temperature in the literature describing DM-baryon interactions \cite[e.g.]{Dvorkin2014, Boddy2018}; however, to avoid confusion with the 21-cm brightness temperature $\delta T_{\rb}$, we instead use $T_{\mathrm{K}}$, consistent with 21-cm literature.}. For Coulomb-like interactions, we consider one target baryon species with $\rho_{\mathrm{t}}=\rho_{\mathrm{b}}$ and a target mass is the mean molecular weight of baryons given by
\begin{equation}
    \mu_{\mathrm{b}}=\frac{m_{\mathrm{H}}}{1+(m_{\mathrm{H}}/m_{\mathrm{He}}-1)Y_{\mathrm{He}}+(1-Y_{\mathrm{He}})x_{\mathrm{e}}}
\end{equation}
where $m_{\mathrm{H}}$ is the mass of neutral hydrogen, $m_{\mathrm{He}}$ is the average mass of neutral helium, $Y_{\mathrm{He}}$ is the primordial helium abundance, and $x_{e}$ is the electron fraction. For millicharged DM interactions, the scattering targets are free protons and free electrons; therefore, $\rho_{\mathrm{t}}$ will depend on the ionized fraction. The factor in the parentheses in Eq.~\eqref{Rchi} captures the velocity dependence of the interaction, taking into account thermal velocities and the effects from the relative bulk velocity. Here, the root-mean-square (rms) DM--baryon bulk relative velocity is defined as \cite{Tseliakhovich2010}
\begin{equation}
    V_{\mathrm{\chi b}}^{2} = \int \frac{dk}{k}\Delta_{\xi} \left(\frac{\theta_{\rb}-\theta_{\chi}}{k^{2}}\right)^{2},
\end{equation}
where $\Delta_{\xi}\simeq 2.4 \times 10^{-9}$ is the primordial curvature variance per log wavenumber $k$.
Given that the rms velocity is an integral over $k$, the velocity dependence of the interaction introduces mixing between different $k$ modes and breaks the linearity of the Boltzmann system of equations. An iterative procedure that sidesteps this issue was proposed in Ref.~\cite{Boddy2018}; however, this method is significantly more computationally expensive. We thus adopt the approximation introduced in Refs.~\cite{Tseliakhovich2010, Dvorkin2014} and represented by Eq.~\eqref{Rchi}, where the thermal contribution $u_\mathrm{\chi \mathrm{t}}\equiv \sqrt{T\chi/m_\chi +T_\mathrm{b}/m_\mathrm{t}}$ and the correction capturing bulk relative velocity $V_{\mathrm{\chi b}}/\sqrt{3}$ are added in quadrature, and $V_{\mathrm{\chi b}}$ is the CDM solution, given by Ref.~\cite{Tseliakhovich2010}. The bulk-velocity correction is constant until $z\sim 10^{3}$ and later decays as $V_{\mathrm{\chi b}} \propto z$ for $z\lesssim 10^{3}$, as shown in Fig.~\ref{fig:Rchi}. We adopt this approximation and solve the linear Boltzmann equations independently for each $k$.\footnote{We have checked this assumption explicitly, finding that the differences in power spectra calculated via the iterative and approximate approaches are minimal, and do not significantly affect our results, consistent with Ref.~\cite{Boddy2018}.}

In Fig.~\ref{fig:Rchi}, we illustrate the momentum-transfer rate for Coulomb-like DM--baryon interactions (where DM scatters with ions and neutral targets), normalized to the comoving Hubble expansion rate $aH$. For $z \gtrsim 10^{4}$, the thermal velocities are much larger than the bulk relative velocity, and the Universe is radiation dominated, so the momentum-transfer rate increases as the temperature of the Universe decreases (to the right of the blue circle). At $z\sim 10^{4}$, the thermal velocities drop below the relative bulk velocity and the Universe transitions to matter domination, leading to a decreasing momentum-transfer rate as a function of time. After recombination (denoted by the red circle), the baryons kinetically decouple from photons, and fall into the potential wells formed by DM; as a result, the relative bulk velocity begins to decrease as $z$ decreases, and the momentum transfer rate once again increases as a function of time.\footnote{Note that this increase in momentum-transfer rate at late times is unique to the interacting models with a power-law index of $n=-4$ in the momentum transfer cross section \cite{KBVG2018, Dvorkin:2019zdi,Dvorkin:2020xga}.} As the DM and baryon fluids become more coupled at late times, the bulk relative velocity dissipates further, increasing the interaction strength, and coupling the fluids even more strongly. 

\subsubsection{Linear Matter Power Spectrum}

We set the initial conditions for the density perturbations according to the cosmological parameters defined in Sec.~\ref{sec:intro} and evolve them via the Boltzmann equations for IDM using a modified version of the publicly available {Cosmic Linear Anisotropy Solving System} code, \verb|CLASS|. Consequently, the predicted value of parameters derived from the matter power spectrum, such as  $\sigma_{8}$, will differ from the value inferred by the \textit{Planck} collaboration assuming a CDM cosmological model.

$v^{-4}$ models lead to enhanced late-time (i.e., post-recombination) DM--baryon scattering (see Fig.~\ref{fig:Rchi}). Consequently, unlike models where the cross section scales as a positive power of the relative velocity, $v^{-4}$ affects the power spectrum on a range of scales, rather than solely imprinting suppression on small scales \cite{Nadler:2019zrb}. The nature of the interactions (i.e., the velocity dependence of the interaction cross section), the value of the cross section coefficient, and the DM mass all determine the $k$-dependence and evolution of this suppression relative to CDM.

\begin{figure*}[t]
    \centering
    \includegraphics[width=\textwidth, keepaspectratio]{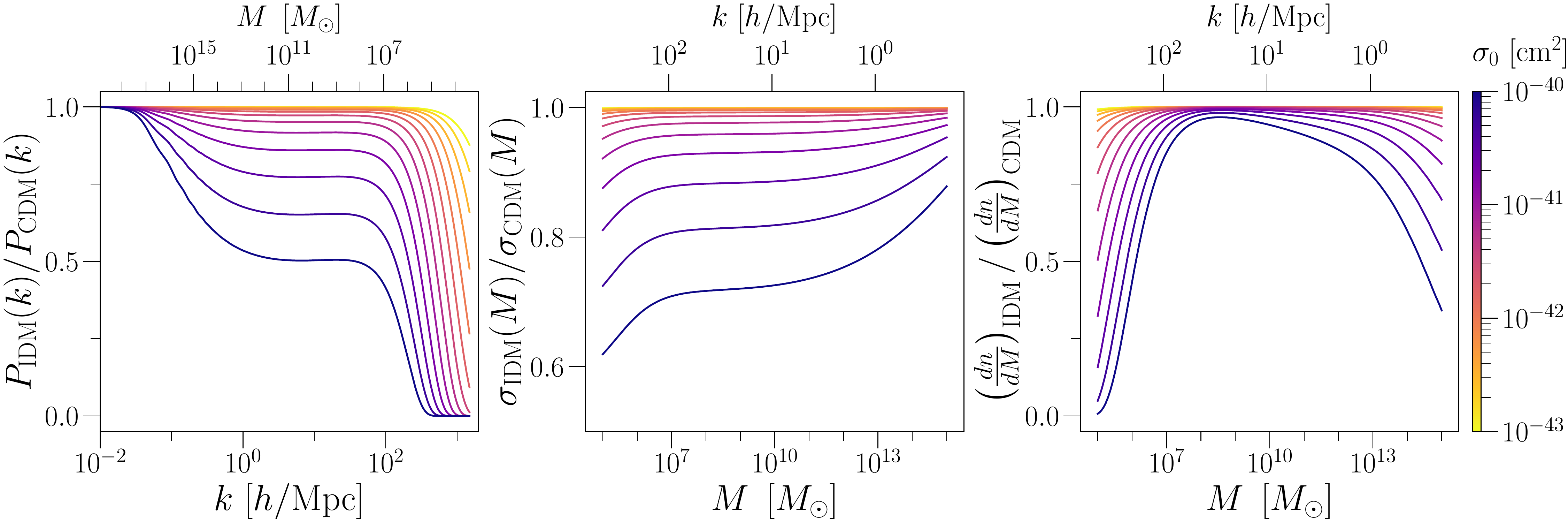}
    \caption{{\textit{Left panel}}: IDM transfer functions: the ratio of matter power spectra in IDM cosmologies with Coulomb-like DM--baryon scattering, $P(k)$, relative to the CDM power spectrum, $P_{\mathrm{CDM}}(k)$, at $z=0$. We fix DM particle mass to $m_{\chi}=1\ \mathrm{MeV}$ and illustrate the transfer function for a range of interaction cross sections, $\sigma_{0}$, indicated by the color bar. Power spectra are obtained using a modified version of the Boltzmann solver \texttt{CLASS}. {\textit{Middle panel}}: The corresponding ratio of the mass variance, $\sigma(M)$, is shown in IDM cosmologies, relative to CDM, at $z=0$, computed using \texttt{Galacticus}. The mass variance is calculated from the power spectrum smoothed via a sharp-$k$ window function (see Section \ref{subsec:nonlinear}). {\textit{Right panel}}: The corresponding ratio of IDM halo mass functions relative to CDM. Halo mass functions are calculated by combining the mass variance with the fitting function from Ref.~\cite{Tinker2008}, using \texttt{Galacticus}. Halo masses $M$ are related to the wavenumber $k$ in linear theory through Eq.~(\ref{eq:k_to_M}).}
    \label{fig:pk_sm_hmf}
\end{figure*}

To illustrate this, the left panel of Fig.~\ref{fig:pk_sm_hmf} shows transfer functions---i.e., ratios of IDM to CDM linear power spectra at $z=0$---for a range of interacting cross sections $\sigma_{0}$, assuming Coulomb-like DM interactions with all baryons. The effects of DM--baryon interactions manifest in two regions of the transfer function. First, starting at $k\sim10^{-2}~h/\mathrm{Mpc}$ (independent of the value of the cross section), there is a loss of power in IDM cosmologies relative to CDM that plateaus at larger $k$, where the height of the plateau scales inversely with $\sigma_{0}$. The plateau is related to the overall momentum transfer, from the time when the corresponding perturbation mode enters the cosmological horizon, until the moment when $R_\chi$ reaches its maximum value and starts to decrease (at $z\sim 10^4$); it is thus controlled by the maximum value of the momentum-transfer rate prior to recombination (Fig.~\ref{fig:Rchi}) and represents a unique feature specific to Coulomb-like IDM. The second feature visible in Fig.~\ref{fig:pk_sm_hmf} is a sharp cutoff at small scales, $k\gtrsim 10^{1}~h/\mathrm{Mpc}$. The cutoff is related to the DM sound horizon scale, and its position depends on the value of the cross section. While the plateau is mostly formed prior to recombination, the small-scale cutoff is driven by late-time collisional damping. However, the small-scale cutoff has little effect on the formation of halos capable of hosting galaxies and only the plateau has a significant impact on the global 21-cm signal, as discussed in the following Section.

For the specific case of millicharged DM, where scattering happens only with ions, the evolution of the momentum transfer rate is identical to other Coulomb-like models until recombination, after which it suffers suppression by the ionized fraction. Since the transfer-function plateau at $k\gtrsim10^{-2}~h/\mathrm{Mpc}$ mostly forms prior to recombination, the structure-formation effects of all $v^{-4}$ models, regardless of the scattering target, are captured by transfer functions shown in  Fig.~\ref{fig:pk_sm_hmf}, for the purposes of the global 21-cm signal.  Ramifications of the small-scale cutoff are only relevant for extremely small scales, beyond the reach of the global signal probes.

\subsubsection{Modified Thermal History}

While the observable suppression of density perturbations occurs mainly prior to recombination, $v^{-4}$ models additionally lead to an exchange of heat between DM and baryons, and a dissipation of their relative bulk velocity, at later times. These processes also affect the global 21-cm signal, by altering the temperature of baryons, as described in Ref.~\cite{Munoz2015}.
To compute the relative bulk velocity ${V}_{\chi \rb}$ after recombination, we evolve the following equation \cite{Munoz2015, Kovetz2018}
\begin{equation}\label{eq:Vchib_evo}
    \dot{V}_{\chi \rb} = -\frac{\dot{a}}{a}V_{\chi \rb} - \left(1+\frac{\rho_{\chi}}{\rho_{\rb}}\right)\sum_{\mathrm{t}}\frac{\rho_{\mathrm{t}}\sigma_{0}F(r_{\mathrm{t}})}{(m_{\chi}+m_{\mathrm{t}})V_{\chi \rb}^{2}},
\end{equation}
where $r_{\mathrm{t}}$, $u_{\chi t}$, and $F(r_{\mathrm{t}})$ are defined by
\begin{equation}
    \begin{split}
        r_{\mathrm{t}}\equiv V_{\chi \rb}/u_{\chi t}, \quad u_{\chi t} \equiv \sqrt{\frac{T_{\chi}}{m_{\chi}}+\frac{T_{\mathrm{K}}}{m_{\mathrm{t}}}} \\
        F(r_{\mathrm{t}}) \equiv \mathrm{Erf}\left(\frac{r_{\mathrm{t}}}{\sqrt{2}}\right)- \sqrt{\frac{2}{\pi}}r_{\mathrm{t}}e^{-r^{2}_{\mathrm{t}}/2},
    \end{split}
\end{equation} 
where $u_{\chi t}$ is the variance of the thermal motion of the two fluids, $r_{t}$ is the ratio of the bulk relative velocity to the thermal velocity, and $F(r_{t})$ is a function related to the drag term that approaches 0 as $r\to 0$ and asymptotes to 1 as $r \to \infty$. The coupled equations for the evolution of the DM and baryon temperatures are given by \cite{Munoz2015, Kovetz2018}
\begin{equation}\label{eq:temp_evo}
    \begin{split}
        \dot{T}_{\chi} = &-2\frac{\dot{a}}{a}T_{\chi} + \sum_{\mathrm{t}}\frac{2}{3}\frac{m_{\chi}\rho_{\mathrm{t}}\sigma_{0}}{u^{3}_{\chi t}(m_{\chi}+m_{\mathrm{t}})^{2}}\\
        & \times \left\{\sqrt{\frac{2}{\pi}}\left(T_{\mathrm{K}}-T_{\chi}\right)e^{-r_{\mathrm{t}}^{2}/2}+m_{\mathrm{t}}\frac{V_{\chi \rb}^{2}}{r_{\mathrm{t}}^{3}}F(r_{\mathrm{t}})\right\}, \\
        \dot{T}_{\mathrm{K}} = &-2\frac{\dot{a}}{a}T_{\mathrm{K}} + \Gamma_{\mathrm{c}}(T_{\mathrm{CMB}}-T_{\mathrm{K}}) \\
        &+ \sum_{\mathrm{t}}\frac{2}{3}\frac{\rho_{\mathrm{t}}\rho_{\chi}\sigma_{0}}{u^{3}_{\chi t}(1+f_{\mathrm{He}}+x_{\mathrm{e}})n_{\mathrm{H}}(m_{\chi}+m_{\mathrm{t}})^{2}}\\
        & \times \left\{\sqrt{\frac{2}{\pi}}\left(T_{\chi}-T_{\mathrm{K}}\right)e^{-r_{\mathrm{t}}^{2}/2}+m_{\chi}\frac{V_{\chi \rb}^{2}}{r_{\mathrm{t}}^{3}}F(r_{\mathrm{t}})\right\}, \\
    \end{split}
\end{equation} 
where the sum is over the target baryon species. For millicharged DM where the sum is performed over ionized species, the target density is a function of the ionized fraction. Therefore, the post-recombination thermal evolution for Coulomb-like interactions and millicharged DM for will differ dramatically until reionization, as the heat exchange in millicharged DM is suppressed by the ionized fraction.

\subsection{Nonlinear Evolution}
\label{subsec:nonlinear}

\subsubsection{Modeling the Halo Mass Function in IDM Cosmologies}

We next model the non-linear gravitational growth of density perturbations into halos capable of forming stars, within a cosmology featuring Coulomb-like DM-baryon interaction. We start with the linear IDM matter power spectrum described in Sec.~\ref{subsec:linear}, and compute the comoving number density of DM halos, referred to as the halo mass function (HMF), using the ePS formalism~\cite{Press:1973iz,Bond:1990iw}. As discussed in Sec.~\ref{subsec:linear}, we find that the Coulomb-like $v^{-4}$ scattering does not affect the linear $P(k)$ on scales of interest at the times when the non-linear growth takes place; we thus assume that the effect of the interactions on the growth rate is negligible at late times and do not attempt to model it in this study. The effects of IDM on HMF are thus modeled through changes in the initial (linear) $P(k)$.

Following the ePS approach, we evaluate the HMF as
\begin{equation}\label{eq:hmf}
    \frac{dn}{dM} = f(\sigma)\frac{\bar{\rho}_{\mathrm{m}}}{M}\frac{d\,  \mathrm{ln}( \sigma^{-1})}{dM},
\end{equation}
where $n$ is the comoving number density of halos of mass $M$, $\bar{\rho}_{\mathrm{m}}$ is the mean matter density, $f(\sigma)$ is a fitting function tuned to match HMFs predicted by N-body simulations, and the filtered mass variance $\sigma(M)$ is computed as an integral of the power spectrum times a window function $W(k|M)$,
\begin{equation}\label{eq:sigmaM}
    \sigma^{2}(M)=\frac{1}{2\pi^{2}}\int_{0}^{\infty} 4\pi k^{2}P(k)W^{2}(k|M)dk.
\end{equation}
A top-hat filter in real space is commonly used to model CDM halo abundances due to the unambiguous relationship between filtering scale $R$ and corresponding halo mass $M$ \cite{Lacey:1994su}. However, this window function is also insensitive to sharp cutoffs in $P(k)$, and consequently poorly predicts the shape of the HMF in cosmologies with suppressed power spectra when compared to N-body simulations~\cite{Benson2012,Schneider:2013ria}. Therefore, we instead adopt the sharp-$k$ filter defined by
\begin{equation}\label{eq:window}
    W(k|M) = \begin{cases} 
    1 & \text{if } k\leq k_{\mathrm{S}}(M)  \\
    0 & \text{if } k > k_{\mathrm{S}}(M)
    \end{cases},
\end{equation}
where
\begin{equation}\label{eq:k_to_M}
    k_{\mathrm{S}}=2.5/R, \quad R = \left(\frac{3M}{4\pi \bar{\rho}}\right)^{1/3}.
\end{equation}
The factor of 2.5 in the conversion between $k_{\mathrm{S}}$ and $R$ was tuned such that the position of the turnover for warm DM (WDM), as a template model for cosmologies with a power-law cutoff, matched the position of the turnover observed in N-body simulations \cite{Benson2012}.
We use the Tinker fitting function \cite{Tinker2008}
\begin{equation}\label{eq:fsigma}
    f(\sigma) = A \left[\left(\frac{\sigma}{b}\right)^{-a}+1\right]e^{-c/\sigma^{2}},
\end{equation}
where $A$ controls the amplitude of $f(\sigma)$, $a$ controls the tilt, $b$ sets the mass scale at which the power law becomes significant, and $c$ determines the high-mass cutoff scale above which halo abundances exponentially decrease. In detail, these parameters are given by
    \begin{equation}
        \begin{split}
            A(z) &= A_{0}(1+z)^{-0.14}, \\
            a(z) &= a_{0}(1+z)^{-0.06}, \\
            b(z) &= b_{0}(1+z)^{-\alpha}, \\
            \log \alpha(\Delta) &= -\left[\frac{0.75}{\log(\Delta/75)}\right]^{1.2},
        \end{split}
    \end{equation}
where we fix the density contrast to $\Delta=500$ in units of the critical density. 
The fitting function in Eq.~(\ref{eq:fsigma}) is widely used and accurately predicts HMFs at low redshifts compared to N-body simulations \citep{2016MNRAS.456.2486D,2022MNRAS.509.6077O}. Although its accuracy is largely unexplored at high redshifts and low halo masses, theoretical uncertainties in the CDM HMF are sub-dominant compared to the effects of our IDM models and the wide range of astrophysical parameters that we test \cite{Mirocha:2020qto}. Thus, we do not vary $f(\sigma)$ in this work.\footnote{We have repeated our this analysis using the Sheth--Tormen HMF fitting function \cite{Sheth2001}, finding that the results are insensitive to the choice of HMF fitting function.}

Using the publicly available \verb|Galacticus| code, we integrate the matter power spectra computed by \verb|CLASS| for Coulomb-like IDM to obtain the mass variance $\sigma(M)$ and corresponding HMF, $\mathrm{\frac{dn}{dM}}$. The output of \verb|Galacticus| is shown for a range of IDM interaction cross sections in Fig.~\ref{fig:pk_sm_hmf}, relative to CDM. We find that $\sigma(M)$ and the HMF are suppressed at every redshift relative to their CDM counterparts. This suppression is most pronounced at halo masses below $\sim$$10^8\,\mathrm{M}_{\mathrm{\odot}}$ and above $\sim$$10^{10}\,\mathrm{M}_{\mathrm{\odot}}$, which we interpret below. At low halo masses, the turnover in the IDM HMF is reminiscent of that in WDM models \cite{Benson2012,Lovell:2013ola,Angulo:2013sza,Schneider:2013ria,Bose:2015mga}; in case of WDM, particles have a significant free-streaming length that produces a cutoff at small scales in $P(k)$ \cite{Viel:2005qj}, similar to that seen in Fig.~\ref{fig:pk_sm_hmf} for IDM. The left panel of Fig.~\ref{fig:hmf_fcoll} shows the redshift evolution of the HMF suppression in IDM, at a few characteristic halo masses.

Finally, we calculate the fraction of mass collapsed into DM halos capable of forming stars, $f_{\mathrm{coll}}$, by integrating over the halo mass function as follows
\begin{equation}
    f_{\mathrm{coll}}(z)=\frac{1}{\bar{\rho}_{m0}(z)}\int_{M_{\mathrm{min}}}^{\infty}M\frac{dn}{dM}dM,
\end{equation}
where $\bar{\rho}_{m0}$ is the mean co-moving mass density and $M_{\mathrm{min}}$ is the minimum halo mass capable of hosting a galaxy; we describe our assumptions regarding $M_{\mathrm{min}}$ in Sec.~\ref{sec:21cm}. The right panel of Fig.~\ref{fig:hmf_fcoll} shows the ratio of the collapsed fraction in our Coulomb-like DM--baryon scattering models relative to CDM. 

\subsubsection{Interpretation}

\begin{figure*}[t]
    \centering
    \includegraphics[width=\linewidth, keepaspectratio]{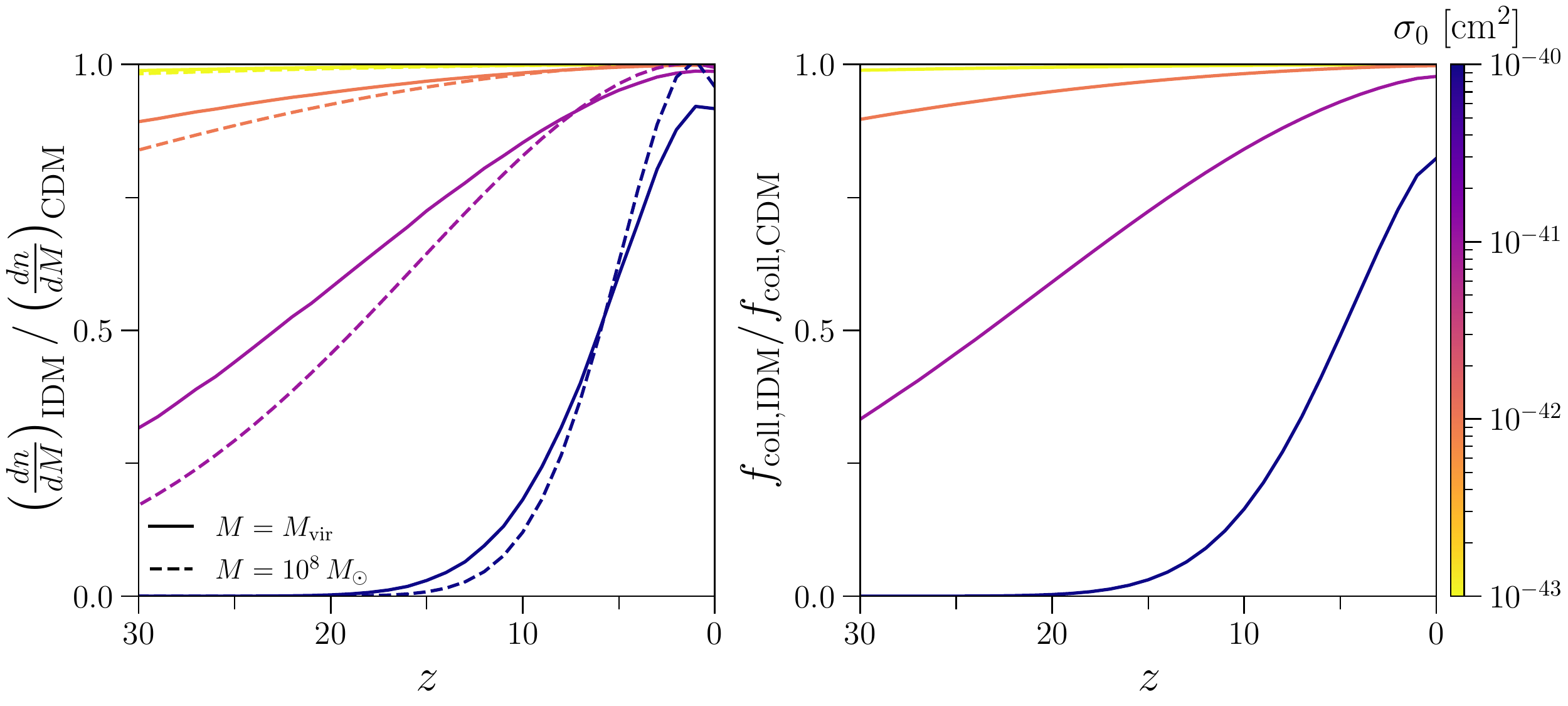}
    \caption{{\textit{Left panel}}: The ratio of the halo mass function (HMF) $\frac{dn}{dM}$ in a Coulomb-like DM--baryon scattering model, relative to CDM, evaluated at the specified halo mass, is shown as a function of redshift, for a range of interacting cross sections (shown in the color bar on the right). DM particle mass is set to $m_{\chi}=1\ \mathrm{MeV}$. The solid line follows the evolution of the HMF at the minimum virial mass corresponding to a virial temperature $T_{\mathrm{vir}}=500\, \mathrm{K}$, while the dashed line follows the evolution of the HMF at a fixed halo mass of $M=10^{8}\, \mathrm{M}_{\odot}$.  {\textit{Right panel}}: The corresponding ratio of the IDM collapsed fraction $f_{\mathrm{coll}}$ relative to CDM, as a function of redshift. The collapsed fraction is calculated by integrating the HMF above the minimum virial mass, set by the virial temperature $T_{\mathrm{vir}}=500\, \mathrm{K}$. Given the greater abundance of smaller mass halos, the collapsed fraction strongly depends on the HMF value at the minimum virial mass.}
    \label{fig:hmf_fcoll}
\end{figure*}

The ratios of the IDM HMFs and collapsed fractions relative to CDM shown in Figs.~\ref{fig:pk_sm_hmf}--\ref{fig:hmf_fcoll} have several interesting features that are relevant for our 21-cm analysis. To develop intuition for these effects, we describe their origins both quantitatively, in the context of the ePS calculations, and qualitatively, in terms of their impact on halo formation.

Quantitatively, for sufficiently large halo masses---i.e., $M_{\mathrm{halo}} \gg M^{*}$, where $M^{*}$ is the characteristic mass scale above which the HMF cuts off---the HMF is exponentially suppressed as a function of the mass variance (see Eq.~(\ref{eq:fsigma})). Thus, the suppression of the mass variance for $10^{8}\,\mathrm{M}_{\mathrm{\odot}}\lesssim M_{\mathrm{halo}}\lesssim 10^{13}\,\mathrm{M}_{\mathrm{\odot}}$ in our IDM models, which results from the plateau in the transfer function at $1~h/\mathrm{Mpc}\lesssim k\lesssim 100~h/\mathrm{Mpc}$, yields a substantial suppression of the HMF for $M_{\mathrm{halo}}\gtrsim 10^8\,\mathrm{M}_{\mathrm{\odot}}$, at $z=0$, as shown in the right panel of Fig.~\ref{fig:pk_sm_hmf}. Heuristically, the scale-independent decrease in power on these scales decreases the abundance of rare density peaks, which correspond to high-mass halos; even a slight decrease in power on these scales thus results in a large suppression of the HMF, because halo abundances fall off exponentially at high masses.  

Meanwhile, at halo masses below $M^{*}$, the small-scale cutoff in $P(k)$ (for $k\gtrsim 10~h/\mathrm{Mpc}$) causes the mass variance to decrease sharply again, suppressing the HMF relative to CDM at the lowest halo masses. The initial lack of power on small scales, produced by collisional damping, inhibits halo formation on the corresponding mass scales, suppressing the amplitude and growth of the HMF. Note that the origin of this low-mass suppression is qualitatively different compared to IDM models in which interactions are efficient well before recombination, suppressing halo abundances on scales that enter the horizon at corresponding times, and then freeze out. In contrast, Coulomb-like interactions become increasingly efficient at later times and therefore do not freeze out. Instead, the small-scale cutoff in the transfer functions for Coulomb-like models is related to the scale of the DM sound horizon, below which the growth of density perturbations is suppressed.

Importantly, because the mass variance and characteristic mass $M^{*}$ increase with time, the turnover in the HMF moves towards larger halo masses with decreasing redshift. Thus, the suppression in the IDM HMF relative to CDM is more severe at early times. This is illustrated by the left panel of Fig.~\ref{fig:hmf_fcoll}; in particular, for sufficiently large cross sections ($\sigma_0\sim 10^{-40}~\mathrm{cm}^2$), almost no halos capable of hosting galaxies exist at $z\gtrsim 15$ in our IDM models. In turn, this is reflected in a suppression of the collapsed fraction of halos capable of hosting galaxies (see the right panel of Fig.~\ref{fig:hmf_fcoll}), which is dominated by the lowest-mass halos above the galaxy formation threshold due to the power-law nature of the HMF.

In summary, the HMF suppression above the characteristic mass scale can affect the global 21-cm signal, because the corresponding halos are large enough to feature molecular-hydrogen cooling and form galaxies early enough to source the Lyman-$\alpha$ background (see Sec.~\ref{sec:21cm}). Meanwhile, the suppression at halo masses below the minimum virial mass, $M\lesssim 10^{7}\,\mathrm{M}_{\odot}$ for the molecular hydrogen cooling threshold, has little impact on the global 21-cm signal, as the corresponding halos are unable to form galaxies. Consequently, since the millicharged and the more general Coulomb-like IDM models differ only in the second cutoff, structure formation for all $v^{-4}$ models can be treated identically, for the purposes of predicting the global 21-cm signal. 

Ideally, our HMF predictions should be calibrated against cosmological IDM simulations to confirm both the quantitative predictions and qualitative reasoning above. However, to our knowledge, no cosmological simulations currently include Coulomb-like interactions, which may necessitate implementing DM--baryon interactions in a hydrodynamic context in addition to modifying initial conditions. We therefore leave a comparison to simulations for future work.

\section{Global 21-cm Signal with Coulomb-like IDM}\label{sec:21cm}

To model the global (sky-averaged) 21-cm signal, we use the publicly available \texttt{ARES} code. Below, we briefly summarize the astrophysical modeling and the general global 21-cm signal modeling, already implemented in \verb|ARES| (Sec.~\ref{sec:ares}). We also describe modifications to \verb|ARES| that were necessary to include the effects of IDM on the thermal history of baryons (Sec.~\ref{sec:idm_ares}). Finally, in Sec.~\ref{sec:putting_it_all_together}, we lay out the procedure for combining predictions for the linear matter power spectrum (using \texttt{CLASS}), HMFs (using \texttt{Galacticus}), and the 21-cm signal (using \verb|ARES|), to self-consistently predict the global signal in IDM cosmologies. 

\subsection{Standard \texttt{ARES} Modeling}
\label{sec:ares}
\subsubsection{Astrophysical Emission} 
To compute the global 21-cm signal from atomic hydrogen in the IGM, \verb|ARES| implements a two-zone model for the IGM, including fully ionized regions (around galaxies) and the bulk IGM, where the global signal originates. The bulk IGM is characterized by the average electron fraction $x_e$ and gas kinetic temperature $T_{\mathrm{K}}$ \cite{Furlanetto2006}, key quantities that affect the global signal. These quantities are in turn controlled by ultraviolet and X-ray emissions from galaxies at redshifts where structure formation and star formation occur. \verb|ARES| models the volume-averaged emissivity of galaxies $\epsilon_{\nu}(z)$ as a function of redshift and wavelength, and uses it as a source term in the cosmological radiative transfer equation \cite[e.g.][]{Haardt1996}, solutions to which yield the mean background intensity $J_{\nu}$. The mean intensity is then integrated to obtain photo-ionization and photo-heating rates. 

To compute the emissivity, in this work we adopt a simple model of star formation, which assumes that the global star formation rate density (SFRD) is proportional to rate at which mass collapses into DM halos, and that the star formation efficiency $f_{\ast}$ is independent of the properties of the halos \cite[see][for generalizations]{Mirocha2017, Park2019}. Therefore, the SFRD $\dot{\rho}_{\ast}$ is proportional to the rate of change in the collapsed mass fraction, above the halo mass threshold for forming galaxies \cite{Barkana2005,Furlanetto2006,Pritchard2012},
\begin{equation}
    \dot{\rho}_{\ast}(z)=\bar{\rho}_{\mathrm{b},0}f_{\ast}\frac{df_{\mathrm{coll}}}{dt}, \label{eq:sfrd}
\end{equation}
where $\bar{\rho}_{\mathrm{b},0}$ is the mean baryon density. The emissivity at a given frequency is then modeled as \cite{Mirocha2014}
\begin{equation}
    \epsilon_{\nu}(z)=f_{i}c_{i}I_{\nu}\dot{\rho}_{\ast}(z), \label{eq:emissivity}
\end{equation}
where $c_{i}$ is the normalization factor that relates $f_{\ast}$ to the energy output in a given emission band $i$ (e.g. Ly-$\alpha$, UV, X-ray), $f_{i}$ is a free parameter that captures the unknown redshift evoluton of $c_{i}$, and $I_{\nu}$ is the spectral energy distribution (SED) of astrophysical sources, such that $\int I_{\nu} d\nu=1$. We describe our choices for these parameters in Sec.~\ref{sec:putting_it_all_together}. 

The emissivity in each band is integrated to find the background intensity $\widehat{J}_{\nu}$ 
\begin{equation}
    \widehat{J}_{\nu}(z) = \frac{c}{4\pi}(1+z)^{2}\int_{z}^{z_{f}}\frac{\hat{\epsilon}_{\nu'}(z')}{H(z')}e^{-\bar{\tau}_{\nu}}dz', \label{eq:intensity}
\end{equation}
where $z_{f}$ is the ``first light redshift" at which astrophysical sources first turn on, $\nu'$ is the emission frequency of a photon emitted at redshift $z'$ observed at frequency $\nu$ and redshift $z$, and $\bar{\tau}_{\nu}$ is the averaged optical depth from $z'$ to $z$. From the background intensity, \verb|ARES| computes the photo-ionization and heating rates in the bulk IGM, using the tables from Ref. \cite{Furlanetto2010} to determine the fraction of secondary photo-electron energy deposited as further ionization and heat. With rate coefficients in hand, the IGM temperature and ionization state are evolved numerically, as in \cite{Mirocha2014}. 

\subsubsection{Global 21-cm Signal}

The observable redshifted 21-cm signal is defined as the difference of the brightness temperature and the background CMB temperature. The signal depends on the ionized fraction $x_e$, the temperature of the bulk IGM $T_{\mathrm{K}}$, and background intensity of Lyman-$\alpha$ $\widehat{J}_{\alpha}$, as \cite{Furlanetto2006}
\begin{equation}
    \begin{split}
        \delta T_{\rb} &= 27 (1-\bar{x}_{i})\left(\frac{1-Y_{p}}{0.76}\right)\left(\frac{\Omega_{\rb}h^{2}}{0.023}\right)\sqrt{\frac{0.15}{\Omega_{\mathrm{m}}h^{2}}\frac{1+z}{10}}\\
        & \times \left(1-\frac{T_{\gamma}}{T_{\mathrm{S}}}\right) \mathrm{mK}
    \end{split}
\end{equation}
where $\bar{x}_{i} = Q + (1 - Q)x_e$ is the mean ionized fraction, $Q$ the volume-filling fraction of fully ionized regions, $Y_{p}$ is the Helium mass fraction, $T_{\mathrm{S}}$ is the spin temperature, and $T_{\gamma}$ is the CMB temperature. The spin temperature $T_{\mathrm{S}}$ quantifies the level populations of the hyperfine singlet and triplet states, and can be written as \cite{Furlanetto2006} 
\begin{equation}
    T^{-1}_{\mathrm{S}} \approx \frac{T_{\gamma}^{-1}+x_{\mathrm{c}}T^{-1}_{\mathrm{K}}+x_{\alpha}T^{-1}_{\alpha}}{1+x_{\mathrm{c}}+x_{\alpha}},
\end{equation}
where $T_{\gamma}$ is the temperature of the CMB, $x_{\mathrm{c}}$ is the collisional coupling coefficient \cite{Zygelman2005}, and $x_{\alpha}$ is the Lyman-$\alpha$ coupling coefficient with a corresponding Ly-$\alpha$ color temperature $T_{\alpha}$. 

Key quantities that affect the spin temperature and that require non-standard modeling in presence of DM-baryon interactions are the gas temperature $T_K$ and the strength of the Lyman-$\alpha$ coupling.  In the next section, we discuss specific modifications we introduced to \texttt{ARES} in order to model these quantities in presence of IDM.

\subsection{Key Modifications to \texttt{ARES} in IDM}\label{sec:idm_ares}

\subsubsection{Gas Kinetic Temperature}

The evolution of the kinetic temperature of the bulk IGM in IDM cosmology is modified with respect to CDM by the presence of DM-baryon heat exchange, and is given by \cite{Mirocha2014,Munoz2015}
\begin{equation}\label{eq:all_heat}
    \frac{3}{2}\frac{d}{dt}\left(\frac{k_{\mathrm{B}}T_{\mathrm{K}}n_{\mathrm{tot}}}{\mu}\right) = \underbrace{\vphantom{\frac{k_{\rb}\dot{Q}_{\rb}n_{\mathrm{tot}}}{\mu}}\phantom{x_{a}}\epsilon_{\mathrm{X}}\phantom{x_{b}}}_\text{Astrophysics} + \,  \underbrace{\frac{k_{\mathrm{B}}\dot{Q}_{\rb}n_{\mathrm{tot}}}{\mu}}_\text{Dark Matter} \, - \underbrace{\phantom{x_{a}}\mathcal{C}\phantom{x_{b}}\vphantom{\frac{k_{\rb}\dot{Q}_{\rb}n_{\mathrm{tot}}}{\mu}}}_\text{CDM} 
\end{equation}
where $k_{\mathrm{B}}$ is the Boltzmann constant, $n_{\mathrm{tot}}$ is the total baryon number density, $\mu$ is the
mean molecular weight, $\epsilon_X$ is the heating rate density from astrophysical sources, $\dot{Q}_{\rb}$ is the heating rate due to the DM--baryon interactions, and $\mathcal{C}$ represents additional heating and cooling terms in a standard CDM cosmology, i.e., Hubble cooling, Compton heating, and collisional ionization cooling  \cite{Seager:1999km, Chluba:2010ca}. 

The ``Dark Matter'' term in Eq.~(\ref{eq:all_heat}) captures the effects of IDM on thermal history of hydrogen in the universe, and has been included in previous studies. We next describe how the effects of altered structure formation in IDM cosmology enter the computation of the 21-cm signal, both through corrections to the Lyman-$\alpha$ coupling at low IGM temperature and through modification of the Lyman-$\alpha$ background.

\subsubsection{Lyman-$\alpha$ Coupling}

The Ly-$\alpha$ emission and the Wouthuysen–Field effect couple the spin temperature to the gas kinetic temperature at Cosmic Dawn \cite{Wouthuysen1952}, driving the depth of the global-signal absorption trough at high redshift. The coupling coefficient is given as
\cite{FurlanettoPritchard2006} 
\begin{equation}
    x_{\alpha} = \frac{16\pi \chi_{\alpha}J_{\infty}}{27 A_{21}}\frac{T_{\ast}}{T_{\gamma}}S_{\alpha},
\end{equation}
where $\chi_{\alpha}=(\pi e^{2}/m_{\mathrm{e}}c)f_{\alpha}$ and $f_{\alpha}$ is the absorption oscillator strength, $A_{21}=2.85\times 10^{-15} \,\mathrm{s^{-1}}$ is the spontaneous emission coefficient of the 21-cm transition, $T_{\ast}=0.068\,\mathrm{K}$ is the energy defect of that transition, and is the $S_{\alpha}$ is a scattering correction factor \cite{Hirata2006,Chuzhoy2006,FurlanettoPritchard2006,Mittal2020}, given by
\begin{equation}
    S_{\alpha} = \int_{-\infty}^{\infty}dx\,\phi(x)\frac{\widehat J_x}{\widehat J_{\infty}}
\end{equation}
where $\phi(x)$ is the normalized Ly-$\alpha$ line profile; $\widehat J_{\infty}$ is the background intensity entering the red wing of the Ly-$\alpha$ line. 

Studies of this scattering correction have to date focused only on the $T_{\mathrm{K}} \gtrsim 1$ K regime, which is appropriate for CDM cosmology. However, we find that in the presence of DM interactions with hydrogen, the IGM can cool down to a fraction of a degree Kelvin, where the behaviour of the Lyman-$\alpha$ coupling has not been studied in detail. In this study, we apply the fitting formula derived in Ref. \cite{Mittal2020}, based on the application of the ``wing approximation" for the line profile \cite{Grachev1989, Chuzhoy2006}, down to arbitrarily low temperatures\footnote{We note that the ``wing approximation'' (and, more broadly, the Fokker-Planck equation from which the background intensity is derived) is expected to break down at extremely low gas temperatures of $T_{\mathrm{K}} \ll 1\ \mathrm{K}$. To properly calculate $J(\nu)$ below these temperatures requires an iterative solution of the integro-differential equations for $J(\nu)$, or a Monte Carlo simulation to sample $J(\nu)$ \cite{Hirata2006}. For simplicity, and as a somewhat conservative approach, we  proceed with the fitting formula of Ref.~\cite{Mittal2020} at all temperatures.}. This can lead to significant changes in the predicted global 21-cm signal, as discussed in Sec.~\ref{sec:edges}.

\subsection{Predicting the 21-cm Signal in IDM Cosmologies}\label{sec:putting_it_all_together}
\label{subsec:full}

By combining the ingredients described above with the structure formation modeling from Sec.~\ref{sec:mc_struct}, we obtain a self-consistent model of the global 21-cm signal in cosmologies with Coulomb-like DM interactions. In particular, we model the 21-cm signal in IDM cosmologies as follows:
\begin{enumerate}
    \item Generate and evolve the linear matter power spectrum $P(k)$ at high redshift ($z=500$), for a chosen set of IDM parameters, using a modified version of \texttt{CLASS} (Sec.~\ref{subsec:linear}). 
    
    \item Compute HMF for the specified IDM cosmology, based on the input $P(k)$ from Step 1, using \texttt{Galacticus} (Sec.~\ref{subsec:nonlinear}).
    
    \item Use the modified HMF from Step 2 to predict the cosmic star formation history and emissivity, including the Lyman-$\alpha$ background and photo-heating of IGM (Sec.~\ref{sec:ares}).
    
    \item Evolve the thermal and ionization histories at later redshifts, corresponding to Cosmic Dawn, using standard \texttt{ARES} (Sec.~\ref{sec:ares}), its IDM modifications (Sec.~\ref{sec:idm_ares}), and the thermal history initial conditions from Step 1.
    
    \item Compute the global signal (Sec.~\ref{sec:putting_it_all_together}).
    
\end{enumerate}

\begin{table*}[t]
    \centering
    \begin{tblr}{|c|c|c|c|c|}
    \hline
    & Parameter & Physical Interpretation & Range / Values \\
    \hline \hline
    \SetCell[r=3]{c} \rotatebox[origin=c]{90}{IDM} &
    $t$ & Target of DM scattering & Baryons (Coulomb-like), Ions (millicharged DM)   \\ 
    \hline
    & $\sigma_{0}$ & DM-target Interaction Strength &  $\left[10^{-43}, 10^{-38}\right]\ \mathrm{cm^{2}}$  \\ 
    \hline
    & $m_{\chi}$ & DM particle mass & $\left[1\, \mathrm{keV}, 10\, \mathrm{GeV} \right]$ \\ 
    \hline \hline 
    \SetCell[r=4]{c} \rotatebox[origin=c]{90}{Astrophysics} &
    $T_{\mathrm{min}}$ & Minimum virial temperature & $500\, \mathrm{K}\, (\mathrm{molec.\, cooling}), \, 10^{4}\, \mathrm{K}\, (\mathrm{atomic\, cooling})$  \\ \hline
    & $f_{\ast}$ & Star formation efficiency & $0.01,\, 0.05$   \\ 
    \hline
    & $f_{\mathrm{X}}$ & X-ray production efficiency & $1,\, 10$  \\ 
    \hline 
    & $f_{\mathrm{esc}}$ & Ionizing escape fraction & $0.1$  \\
    \hline
    \end{tblr}
    \caption{IDM and astrophysical parameters used to calculate the evolution of the global 21-cm signal in our analyses.}
    \label{tab:astro_params}
\end{table*}
This algorithm is illustrated in Fig.~\ref{fig:pipeline}.
In order to predict the global signal, it is necessary to also specify values for the astrophysical parameters that relate the collapsed fraction to the emission in each frequency band. Ideally, the astrophysical and IDM parameters would be simultaneously assessed to determine viable portions of the overall parameter space. However, this would come at a significant computational expense, so here, we choose a set of representative values of astrophysical parameters and leave a simultaneous sampling of the entire parameter space for future work. Our choices for the astrophysical parameters are summarized in Tab.~\ref{tab:astro_params}, and are as follows:

\begin{itemize}
    \item \emph{Minimum virial temperature} $T_{\mathrm{min}}$: We choose two values of $T_{\mathrm{min}}$, $500\, \mathrm{K}$ and $10^{4}\, \mathrm{K}$, which are motivated by the molecular and atomic hydrogen cooling thresholds, respectively \cite[e.g.,][]{Tegmark1997}. A smaller minimum virial temperature leads to a larger $f_{\mathrm{coll}}$ at each redshift, and so is also more sensitive to the suppression of smaller mass halos in IDM.
    
    \item \emph{Star formation efficiency} $f_{\ast}$: We choose a constant star formation efficiency of $f_{\ast}=0.05 \text{ and } 0.01$, for all halo masses, motivated by the peak value derived from fitting galaxy luminosity functions in Ref.~\cite{Mirocha2017} and a value that more closely matches the SFRD predicted by the same model. Both values of $f_{\ast}$ likely overestimate the overall SFRD, leading to a conservative prediction for the delay of the global signal due to IDM, given the degeneracy between $f_{\ast}$ and $f_{\mathrm{coll}}$.\footnote{For the rest-ultraviolet emission of galaxies, we adopt the spectrum of a $1/5$ solar metallicity galaxy forming stars at a constant rate as given by the \texttt{BPASS} version 1 single-star models \cite{Eldridge2009}.}
    
    \item \emph{X-ray efficiency} $f_{\mathrm{X}}$: We focus on two choices: $f_{\mathrm{X}}=1$, which reproduces the local $L_X$/SFR relation \cite{Mineo2012}, and $f_{\mathrm{X}}=10$, which is likely representative of low-metallicity high-$z$ galaxies \cite{Brorby2016, Fragos2013,Fornasini2019,Lehmer2022}. In each $f_X$ case, we adopt an X-ray spectrum representative of high-mass X-ray binary systems, i.e., a hard multi-colour disk spectrum \cite{Mitsuda1984} that assumes $10 \ M_{\odot}$ black hole primaries. We neglect the effects of neutral attenuation within galaxies, which would lead to an even harder X-ray source spectrum and less efficient heating \cite{Mirocha2014,Das2017}.
    
    \item \emph{Ionizing escape fraction} $f_{\rm{esc}}$: We fix the ionizing escape fraction $f_{\rm{esc}}=0.1$ in this study. In reality, scenarios with significant suppression of structure would require an increase in $f_{\rm{esc}}$ in order to remain consistent with constraints on, e.g., the Thomson optical depth, $\tau_e$. We defer a self-consistent exploration of Cosmic Dawn and reionization to future work.
\end{itemize}
We refer the reader back to Sec.~\ref{sec:ares} for details regarding other astrophysical parameters and modeling choices.

\section{Results}\label{sec:edges}

We now describe the results of our 21-cm predictions in IDM cosmologies. We begin with a discussion of the general features of the global 21-cm signal in Coulomb-like and millicharged DM--baryon scattering models in Sec.~\ref{sec:examples}. We then discuss the implications of our predictions in the context of the EDGES measurement in Sec.~\ref{sec:implications}, focusing on Coulomb-like and millicharged models separately.

\subsection{Features of the 21-cm Signal in IDM}\label{sec:examples}

\begin{figure}[t]
    \centering
    \includegraphics[width=\linewidth]{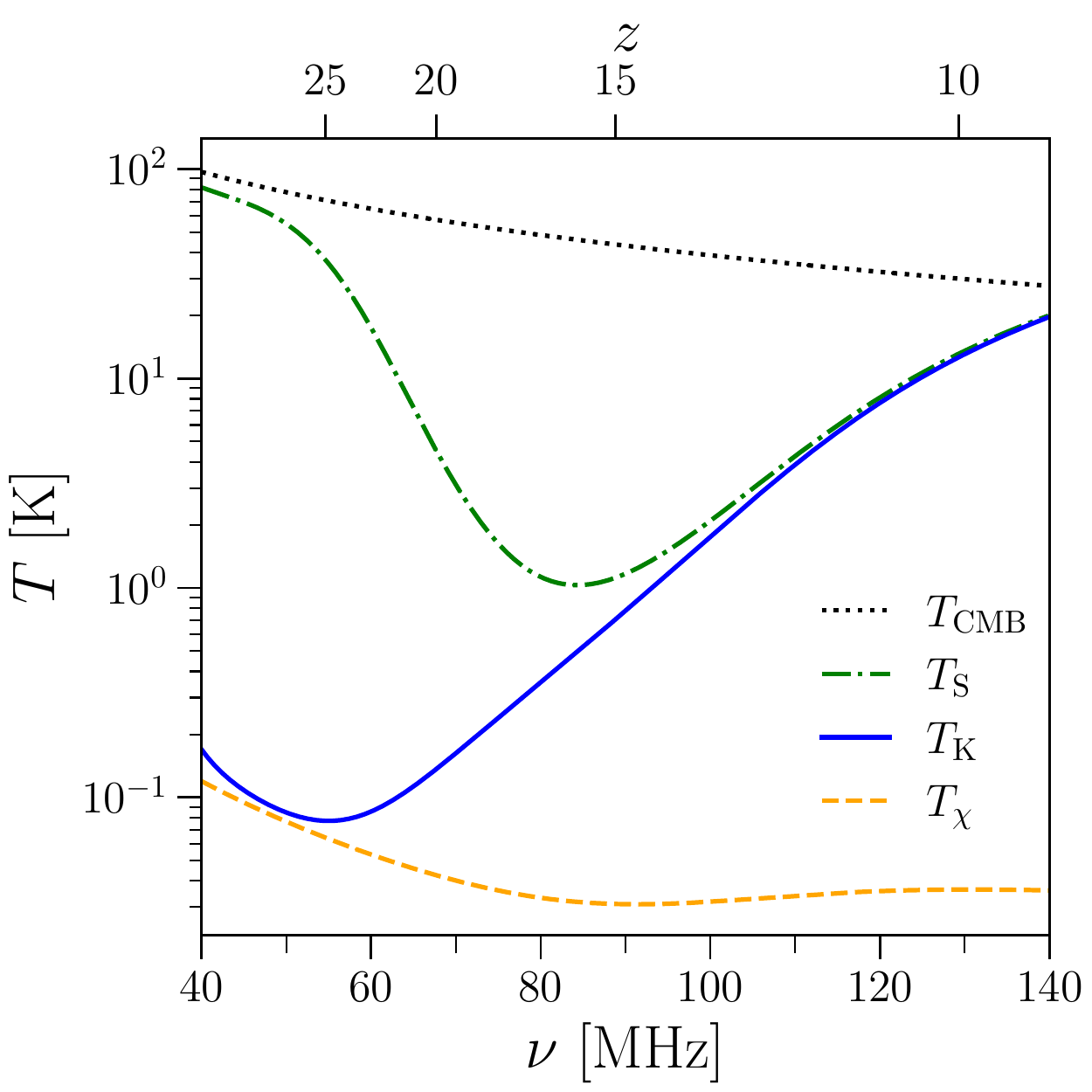}
    \caption{We show the redshift evolution of the CMB temperature, $T_{\mathrm{CMB}}$ (black), 21-cm spin temperature, $T_{\mathrm{S}}$ (green), gas kinetic temperature, $T_{\mathrm{K}}$ (blue), and DM temperature, $T_{\chi}$ (orange), for a Coulomb-like IDM model with a scattering cross section $\sigma_{0}=10^{-41}\ \mathrm{cm^{2}}$ and DM particle mass $m_{\chi}=10\,\mathrm{MeV}$. $T_{\mathrm{S}}$ is initially coupled to $T_{\mathrm{CMB}}$ until the production of Lyman-$\alpha$ photons increases around $\nu \sim 60$ MHz, at which point it couples to the gas temperature $T_{\mathrm{K}}$. The suppression of the formation of structure in an IDM cosmology delays the coupling of $T_{\mathrm{S}}$ to $T_{\mathrm{K}}$, shifting the minima of $T_{\mathrm{S}}$ to later times.}. 
    \label{fig:temperatures}
\end{figure}

In Fig.~\ref{fig:temperatures}, we illustrate the evolution of the spin temperature in an IDM cosmology which, along with the mean ionized fraction, determines the global 21-cm signal. For this illustration, we fix the cross section $\sigma_{0}=10^{-41}\,\mathrm{cm^{2}}$ and mass $m_{\chi}=1\,\mathrm{MeV}$. We see that, at $\nu\lesssim 50\,\mathrm{MHz}$, the spin temperature is coupled to the CMB and the gas temperature is nearly coupled to the DM temperature; this holds for all sufficiently large cross sections, $\sigma_0\gtrsim 10^{-41}~\mathrm{cm}^2$. Thereafter, a sufficient amount of matter has collapsed into halos such that the production of Lyman-$\alpha$ and X-ray photons becomes significant. The emergence of the Lyman-$\alpha$ background ``activates'' the 21-cm signal, driving $T_\mathrm{S}\rightarrow T_\mathrm{K}$ via the Wouthuysen-Field effect while X-rays heat the gas. This hastens the coupling of $T_\mathrm{S}$ to $T_\mathrm{K}$ by increasing the Wouthuysen-Field coupling strength, and eventually drives $T_\mathrm{K}$ toward $T_\mathrm{CMB}$. 
Given that DM is still in thermal contact with baryons, the DM temperature is also raised, slowing the rate at which the gas temperature is heated in response to the production of X-ray photons.

\begin{figure*}[t]
    \centering
    \includegraphics[width=\linewidth]{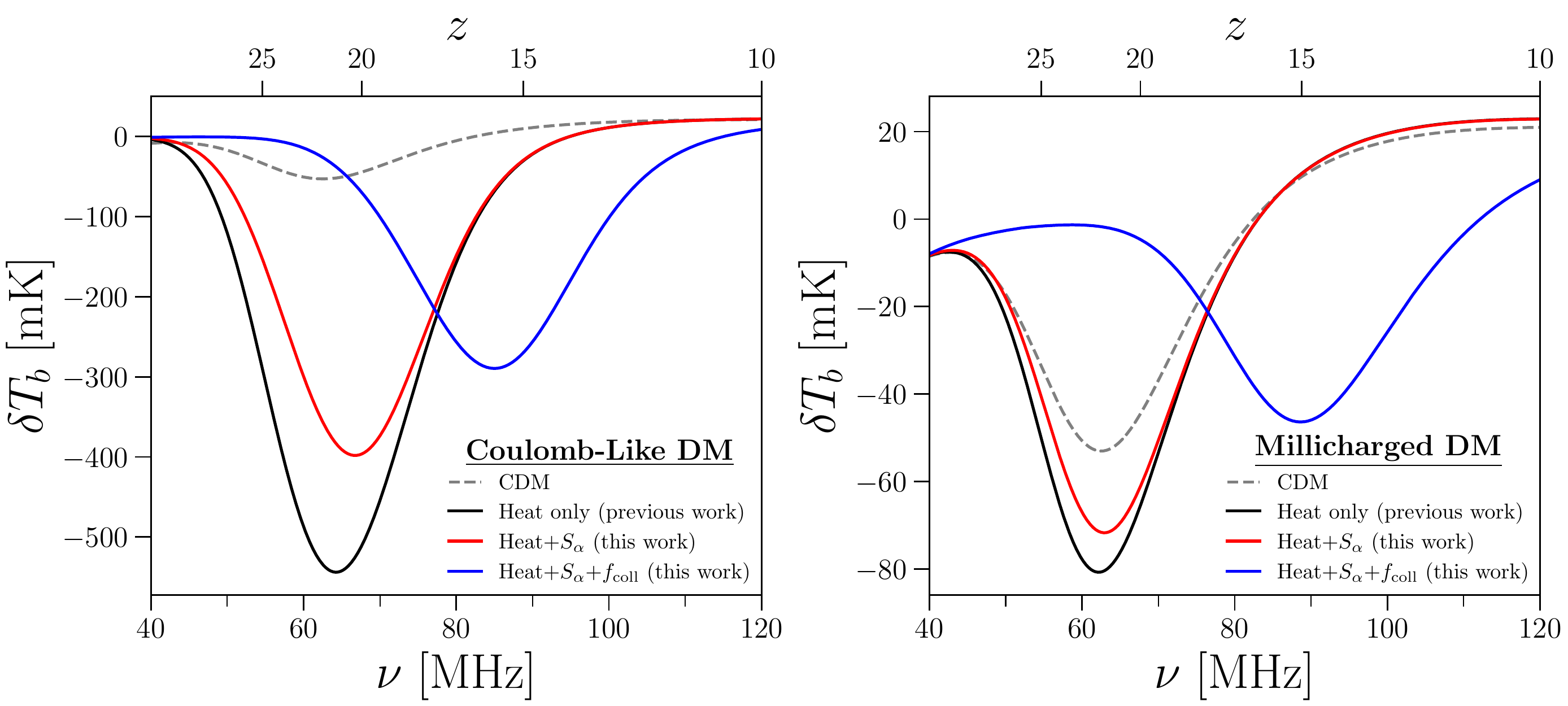}
    \caption{The global 21-cm absorption signal from Cosmic Dawn, in presence of Coulomb-like DM--baryon scattering (left) and in presence of  millicharged DM (right). The CDM case is shown as a dashed line. Different solid lines correspond to 21-cm signal modeling that includes: (1) only the effects of the heat exchange, as studied in previous work (black), (2) heat exchange and the corrected modeling of the Lyman-$\alpha$ coupling at gas kinetic temperatures below $\sim 1\,\mathrm{K}$, presented in this work (red), and (3) the suppression of structure, in addition to the heat exchange and the corrected Lyman-$\alpha$ coupling---i.e., the full calculation presented in this work (blue). These predictions are computed for a fixed DM--baryon scattering cross section $\sigma_{0}=10^{-40}\,\mathrm{cm^{2}}$ and DM particle mass $m_{\chi}=4.6\times 10^{-1}\,\mathrm{GeV}$. 
    All other cosmological parameters are fixed to the \textit{Planck} best-fit values, astrophysical parameters are fixed to $T_{\mathrm{min}}=500\,\mathrm{K}$, $f_{\ast}=10^{-2}$, and $f_{X}=10$. The Lyman-$\alpha$ coupling correction significantly modifies the amplitude of the signal, while the suppressed structure formation significantly delays the signal. Neglecting these effects leads to an overestimate in the amplitude of the signal by $\sim 300\, \mathrm{mK}$ and a shift in the timing of the signal by $\sim 30\, \mathrm{MHz}$.}
    \label{fig:model_comp}
\end{figure*}

Next, in Fig.~\ref{fig:model_comp} we compare previous calculations of the global signal (or, the brightness temperature $\delta T_{\rb}$) for the CDM case (dashed line) to that for Coulomb-like (panel on the left) and millicharged DM (panel on the right) models; for the IDM models, we show the previous IDM heat-exchange-only calculations (black line) and successively include the scattering correction $S_{\alpha}$ (red) and the collapsed fraction calculated from the IDM HMF (blue line), which is our fully-consistent model of the global signal in IDM. In this Figure, we see how the signal changes as individual effects on thermal history and structure-formation delay are included in the calculation. In particular, the scattering correction significantly weakens the coupling for cold IGM, decreasing the depth of the absorption trough. Meanwhile, the reduction in the collapsed fraction delays the onset of the trough, for large values of the interaction cross section. This general behavior holds true regardless of the choice of astrophysical parameters. 

\begin{figure*}[t]
    \centering
    \includegraphics[width=\linewidth]{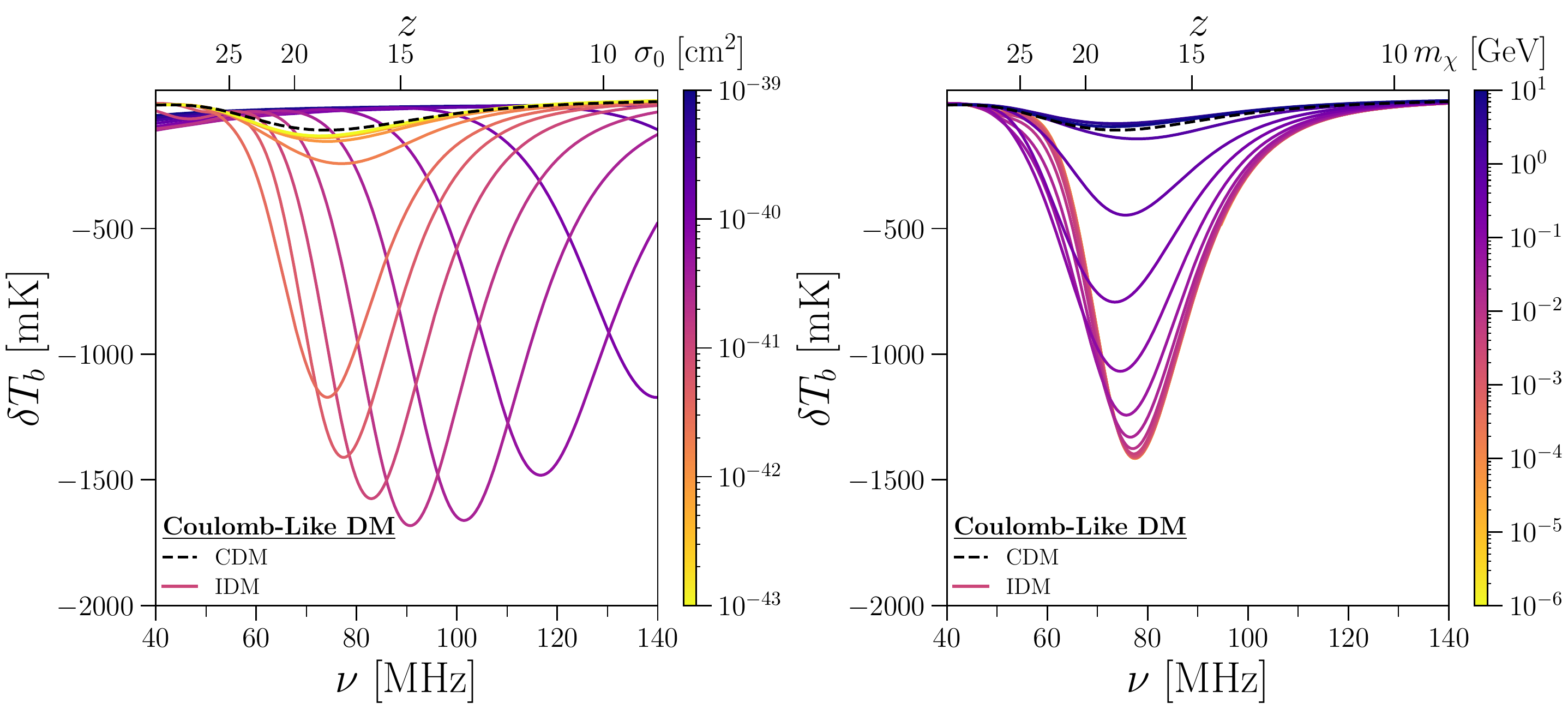}
    \caption{\emph{Left panel}: 21-cm brightness temperature $\delta T_{\rb}$ plotted over time for different values of the interacting cross section $\sigma_{0}$ for Coulomb-like DM with a fixed DM particle mass $m_{\chi}=1\,\mathrm{MeV}$ and astrophysical parameters $T_{\mathrm{min}}=500\,\mathrm{K},\ f_{\ast}= 10^{-2},\text{ and}\ f_{\mathrm{X}}=1$. As the interaction strength increases, the amplitude of the absorption trough grows, and the signal shifts to later times. \emph{Right panel}: The corresponding global signal for a fixed cross section $\sigma_{0}=5.6\times 10^{-42}\,\mathrm{cm^{2}}$, instead varying the DM particle mass $m_{\chi}$ with the same fixed astrophysical parameters as the left panel. Below $m_{\chi}=1\,\mathrm{MeV}$, there is minimal dependence on the DM mass. However, the heat and momentum transfer rate is suppressed as the DM particle mass increases above the proton mass, in a manner that is nearly degenerate with decreasing the interaction strength.}
    \label{fig:gs_sm}
\end{figure*}

Finally, Fig.~\ref{fig:gs_sm} illustrates the effects of varying IDM parameters on the global signal, for a fully-consistent signal modeling (including the heat exchange, effects on structure, and the Lyman-$\alpha$ coupling correction). We note that the depth and the timing of the trough both depend on the interaction cross section (illustrated in the panel on the left) and on the DM particle mass (illustrated in the panel on the right). In particular, as the interaction strength increases, the gas cools more efficiently until it reaches thermal equilibrium with DM, prior to Cosmic Dawn, leading to a deeper absorption trough. At large enough interaction strength, however, the gas temperature dramatically decreases and the Lyman-$\alpha$ coupling becomes inefficient; for the largest cross sections shown in this Figure, the trough is thus less deep. Meanwhile, the delay of the signal becomes more pronounced at higher cross sections, because of the suppression in the collapsed fraction (Fig.~\ref{fig:hmf_fcoll}). 

On the other hand, variations in the IDM mass depend on the mass of the target. When $m_{\chi}\gtrsim m_{\mathrm{t}}$, the heat and momentum transfer rates are inversely proportional to the IDM mass, so increasing the IDM mass is equivalent to decreasing the interaction strength. However, when $m_{\chi} \ll m_{\mathrm{t}}$, the momentum transfer rate is independent of the IDM mass, while the heat transfer rate varies inversely with the IDM mass. This results in a colder DM temperature post-recombination and a delayed late-time heat exchange between DM and baryons. If the interaction strength is strong enough to thermally couple DM and baryons, the global signal will deepen with decreasing IDM mass.


\subsection{Implications in the Context of EDGES}\label{sec:implications}

In order to illustrate how our results would translate to constraints on IDM, under the interpretation of the EDGES measurement as a global signal, we define a ``window of consistency,'' which corresponds to the $99\%$ confidence interval on the amplitude of the absorption-trough minimum inferred from the EDGES data \cite{Bowman2018}; the width of the window corresponds to the entire width of the trough, as measured by EDGES, and the window is shown as a dashed rectangle in Fig.~\ref{fig:allowed_gs}. For a fixed set of astrophysical and IDM parameters, if a minimum of the predicted global signal lies in this window and occurs between $60~\mathrm{MHz} \leq \nu_{\mathrm{min}} \leq 100~\mathrm{MHz}$ and the amplitude of the minimum lies between $-1000~\mathrm{mK}\leq \delta T_{\mathrm{b,min}} \leq -300~\mathrm{mK}$, we label the predicted global signal as ``consistent'' with the EDGES measurement. Meanwhile, any curve without an absorption trough in this region is labeled ``inconsistent.''

These consistent/inconsistent labels act as generous proxies for the bounds we would obtain from sampling the full astrophysical-plus-IDM parameter space in a likelihood-based data analysis. We expect that our method for evaluating individual 21-cm signals is conservative, and significantly stronger bounds on IDM would be obtained from a full sampling analysis. This is particularly relevant in the context of the EDGES absorption trough, which exhibits a flattened peak and a sharp rise into and out of absorption, which are difficult to reproduce in detail regardless of the mechanism invoked to explain the EDGES trough's amplitude \citep{Kaurov2018,Mirocha2019,Mebane2020,Chatterjee2020,Mittal:2022twx}. We leave this detailed and computationally expensive approach for future work.

We further define a region of IDM parameter space that is consistent with the EDGES measurement if the criterion described above is met for any set of values of the astrophysical parameters we consider. This procedure is again conservative, in the sense that extreme variations in astrophysical parameters may be required to enable the consistency we report for some IDM parameters. We note that, according to this definition, our baseline CDM model, roughly equivalent to an IDM model with a small cross section $\sigma_{0}\lesssim 10^{-43}$, is inconsistent with the EDGES measurement, taken at face value, for all variations of astrophysical parameters we consider.\footnote{Given that the chosen range of $f_{\ast}$ and $f_{\mathrm{X}}$ were based on the SFRD inferred from an extrapolation of the UVLF, this is consistent with past results that observed a discrepancy between the SFRD necessary to produce an EDEGS-like amplitude in CDM and the SFRD inferred from the UVLF \cite{Mirocha2019}, albeit under the assumption of a different star formation model.}

\subsubsection{Coulomb-like Interactions}

We first calculate the global signal for Coulomb-like DM--baryon interactions over a uniform log-space grid of cross sections, $\sigma_{0}=\left[10^{-43}, 10^{-38}\right]\, \mathrm{cm^{2}}$, DM particle masses $m_{\chi}= \left[1\, \mathrm{keV}, 10\, \mathrm{GeV} \right]$, and permutations of the astrophysical parameters $T_{\mathrm{min}}$, $f_{\ast}$, and $f_{\mathrm{X}}$ according to the ranges in Tab.~\ref{tab:astro_params}.\footnote{Assuming thermal production, the low-mass end of our IDM parameter space is WDM, constrained by the bounds on the number of relativistic degrees of freedom and by small-scale structure formation. We do not model relativistic effects in this work, and we only use the EDGES signal to constrain the IDM models, and in that sense, the low-mass exclusion regions we show in this work are conservative.}

\begin{figure*}[t]
    \centering
    \includegraphics[width=\linewidth]{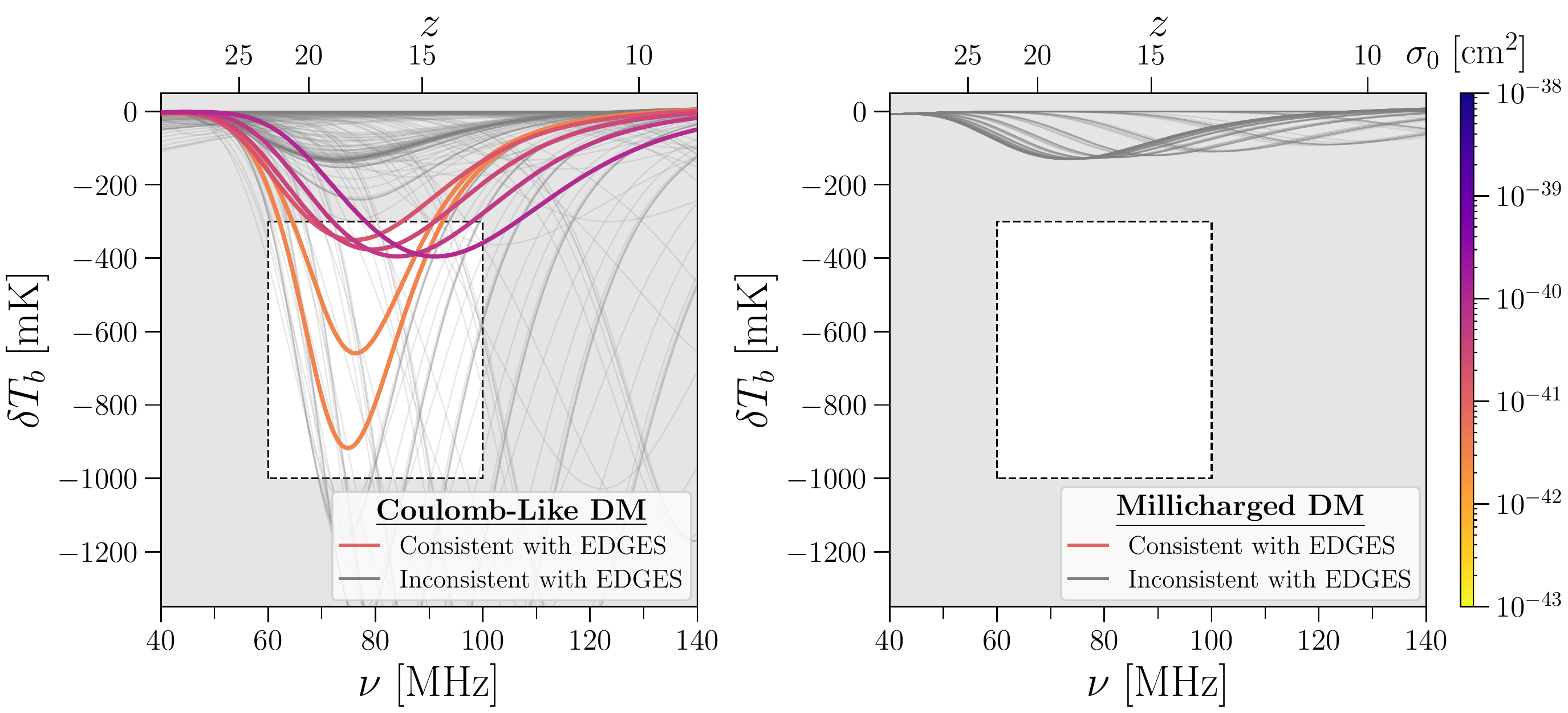}
    \caption{21-cm brightness temperature vs.\ frequency for Coulomb-like DM (left) and millicharged DM (right), for a fixed set of astrophysical parameters $T_{\mathrm{min}}=500\, \mathrm{K}$, $f_{\ast}=10^{-2}$, $f_{X}=1$ and varying IDM cross sections and DM particle masses. Signals with amplitudes that are too small, too large, or too late, relative to a broad frequency and brightness-temperature region where the EDGES collaboration reports measurement of the absorption trough (labeled as the dashed box) are shaded in gray and labeled ``inconsistent" with EDGES; ``consistent'' signals have the trough minima that lie within the consistency box, and are shown in color. For millicharged DM, any interaction strength sufficiently enhances the global signal to match the EDGES depth, simultaneously delays the signal beyond the observed frequency range. Thus, millicharge cannot produce a signal consistent with the EDGES measurement, for any cross section or mass.}
    \label{fig:allowed_gs}
\end{figure*}
\begin{figure}[t]
    \centering
    \includegraphics[width=\linewidth]{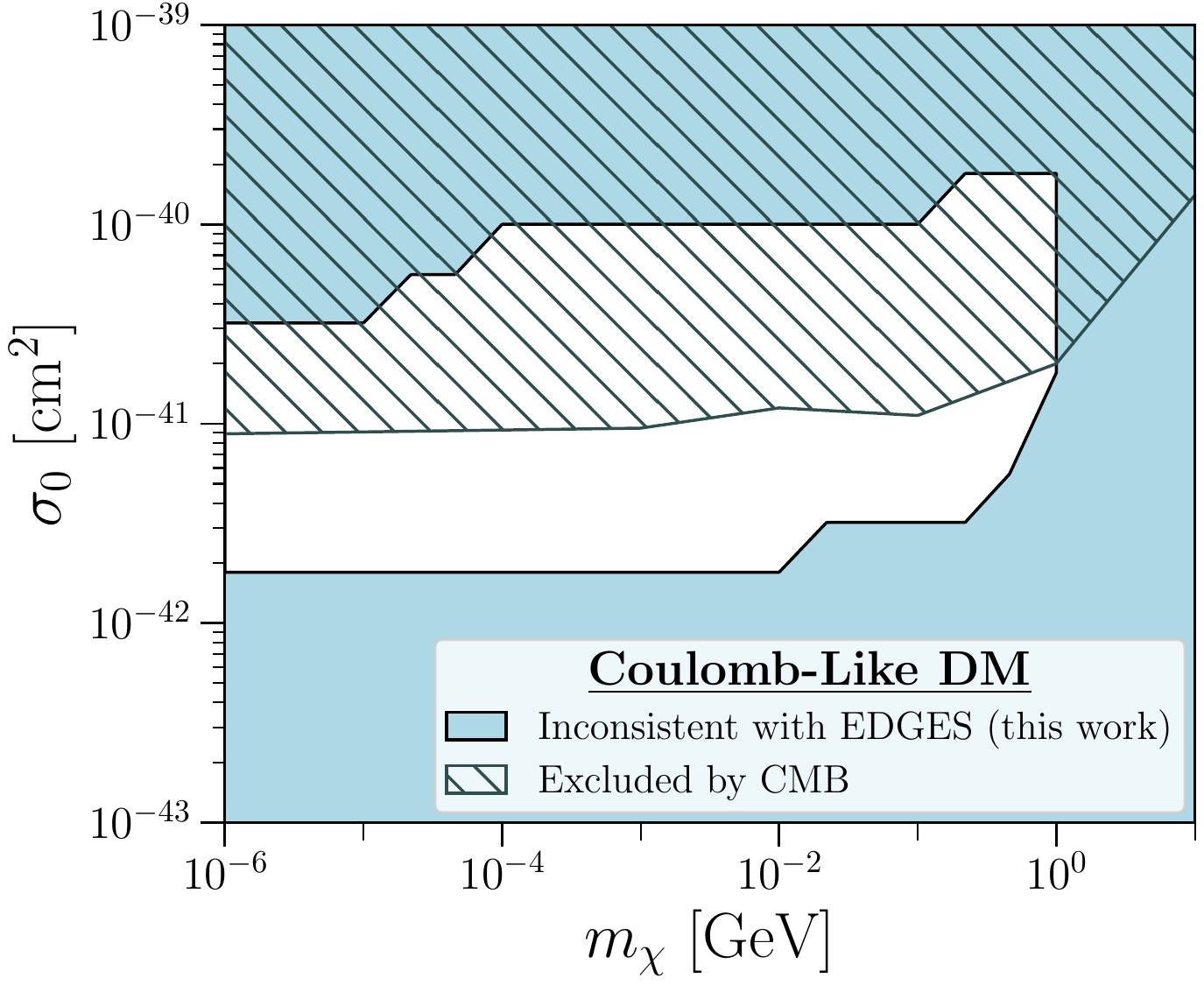}
    \caption{The region of Coulomb-like IDM parameter space inconsistent with the EDGES data, based on the criterion shown in Fig.~\ref{fig:allowed_gs}, is shaded in blue. For comparison, the parameter space inconsistent with the \textit{Planck} measurement of the CMB anisotropy is hatched \cite{Nguyen2021}. We obtain a minimum cross section nearly an order of magnitude larger than in Ref.~\cite{Barkana:2018qrx}, due to the scattering correction and the inclusion of sources of astrophysical heating. Only the white region remains the viable parameter space for EDGES.}
    \label{fig:allowed_sm}
\end{figure}
    
In order to summarize these effects for the entire interacting cross section and DM particle mass parameter space, we apply the method described above to evaluate whether the brightness temperature curve we calculate for each IDM model is consistent with the EDGES measurement. The resulting ``consistent'' and ``inconsistent'' 21-cm predictions are shown in Figure \ref{fig:allowed_gs}, where signals are colored and labeled ``consistent" if the minimum of the curve falls within the broad frequency and amplitude region where EDGES reports the signal, and ``inconsistent" if the minimum occurs outside the broad EDGES window. If the depth of the global signal is too small relative to the EDGES measurement, the interaction strength is too weak to account for the reported anomaly. Meanwhile, some global signal minima are either deeper or occur later than the EDGES measurement, due to DM--baryon interactions that are too strong and/or delay the onset of the 21-cm signal too drastically. The corresponding IDM cross sections are thus too large to explain the data.

We show the corresponding IDM parameter space that is ruled out by the EDGES measurement in Fig.~\ref{fig:allowed_sm}, where the colored region is ``inconsistent.'' For the astrophysical parameters we consider, the lower edge of the white ``consistent'' region defines the minimum cross section necessary to cool the gas sufficiently to produce a deep enough absorption trough; the cross sections above the upper edge delay the global signal beyond the allowed window. Relative to the previous analysis in Ref.~\cite{Barkana:2018qrx}, our results indicate a significantly smaller viable parameter space. When combined with the upper bound on the interaction cross section from the CMB \cite{Nguyen2021}, shown as the hatched region, only a small range of interaction strengths are allowed. 

\subsubsection{Millicharged Interaction}

We repeat a similar analysis for millicharged DM that only scatters with ionized targets. In spite of the generous criterion we choose for consistency of a signal with the EDGES measurement, we find that it is \textit{impossible} to produce the EDGES signal with millicharged DM. Any signal that is sufficiently deep to match the EDGES absorption trough is also delayed far beyond the desired frequency range. The right panel of Fig.~\ref{fig:model_comp} illustrates this response of the global signal as we include the effects of the suppression of the formation of structure. 

Compared to Coulomb-like scattering, the significant delay in the 21-cm signal for millicharged DM models that produce EDGES-like absorption trough depths results from the reduction in the efficiency of heat exchange. In particular, heat exchange between DM and baryons due to millicharged DM interactions scales with the ionized fraction, implying that a significantly stronger cross section than in the Coulomb-like scattering model is necessary to cool the gas temperature prior to reionization sufficiently to match the amplitude of the EDGES measurement. We note that, at fixed cross section, there are only marginal differences in momentum exchange---and thus in the transfer functions---between the Coulomb-like and millicharged DM models.\footnote{The second cutoff in the millicharged DM transfer function reaches a suppression comparable to that in the Coulomb-like scattering case at very late times, after reionization. Nonetheless, as emphasized before, the second cutoff does not significantly affect our analysis for either model.} Thus, the large millicharged DM cross sections necessary to produce a deep absorption trough drastically delay structure formation.

Ref.~\cite{Barkana:2018qrx} and related studies analyzed 21-cm signals with millicharged DM without accounting for suppressed structure formation or additional sources of astrophysical heating, and found that the minimum millicharged DM cross section necessary to sufficiently deepen the global signal is $\sigma_{0}\sim 10^{-39}~\mathrm{cm}^2$. This value is an order of magnitude larger than the maximum consistent cross section we derive for Coulomb-like scattering, which is driven by the timing of the global signal. Because the impact on structure formation is similar for Coulomb-like and millicharged DM models, we conclude that IDM models with $\sigma_{0}\gtrsim 10^{-40}~\mathrm{cm}^2$ delay the global signal far beyond that indicated by the EDGES measurement, regardless of the scattering target. 

In summary, independent of the constraints placed on millicharged DM by the CMB, scenarios where all of the DM is millicharged cannot self-consistently predict the depth and timing of the EDGES-reported global 21-cm signal within the astrophysical parameter space we consider. This conclusion does not rely on the results of CMB analyses, which find that millicharged DM can only explain EDGES if it constitutes a sub-percent fraction of the total dark matter content \cite{Kovetz2018}. 

\section{Discussion}
\label{sec:discussion}

Our results demonstrate how the global 21-cm signal responds to the presence of Coulomb-like and millicharged DM--baryon interactions over a wide range of relevant astrophysical parameters. Here, we discuss how our findings depend on astrophysical assumptions (Sec.~\ref{sec:astro_dependence}), what their implications are for future measurements (Sec.~\ref{sec:21cm_future}), and how Coulomb-like DM--baryon interactions impact structure formation beyond 21-cm cosmology (Sec.~\ref{sec:other_effects}).

\subsection{Dependence on Astrophysical Assumptions}\label{sec:astro_dependence}

The minimum cross section we find to be consistent with the EDGES measurement, for Coulomb-like interaction, depends on the assumed astrophysical parameters. For example, an increase in the scattering correction $S_{\alpha}$ and/or an increase in the background Ly-$\alpha$ flux $\widehat{J}_{\alpha}$ deepens the absorption trough at a fixed DM-baryon scattering cross section; this, in turn, would lead to smaller cross section being consistent with EDGES, and move the lower boundary of the white region in Fig.~\ref{fig:allowed_sm} towards smaller cross section values. Conversely, new sources of heating, or an enhancement in the production of X-ray photons, can further heat the IGM gas, necessitating a more efficient DM-baryon scattering to cool the gas to temperatures to values consistent with EDGES; the increase in $f_X$ would thus drive the lower bound in Fig.~\ref{fig:allowed_sm} towards higher cross-section values. 

In contrast, the upper bound on the cross section in Fig.~\ref{fig:allowed_sm} mainly depends on the timing of the global signal, which is most sensitive to the steepness of the SFRD and efficiency of Ly-$\alpha$ and X-ray photon production. At a fixed interaction cross section, enhancing the emission in all bands by increasing the star formation efficiency $f_{\ast}$, or by decreasing the minimum virial temperature $T_{\mathrm{min}}$, shifts the signal to an earlier time (higher redshift). To compensate for this and preserve consistency of the signal with EDGES, the upper bound of the white region in Fig.~\ref{fig:allowed_gs} would shift toward higher cross section values. For this reason, our assumption of a constant star formation efficiency and the resulting higher SFRD (compared to a more realistic double power law model at the same peak efficiency) is conservative, for the purpose of deriving the bounds in Fig.~\ref{fig:allowed_gs}, as it allows most of the IDM parameter space. 

To ensure our bounds in Fig.~\ref{fig:allowed_gs} are robust to these parameter degeneracies, our window of consistency allows for rather extreme variations in astrophysical parameters, according to the discussion in Sec.~\ref{sec:putting_it_all_together}. If \textit{any} combination of the astrophysical parameters we consider results in a global signal positioned within the frequency and the brightness-temperature interval consistent with the EDGES-inferred absorption-trough minimum, the corresponding IDM model is labeled as ``consistent.'' In future work, we plan to simultaneously sample and constrain astrophysical and IDM parameters in a likelihood analysis that accounts for the depth, timing, and shape of the global 21-cm measurements.

\subsection{Implications for Future 21-cm Measurements}\label{sec:21cm_future}

Our analysis in Sec.~\ref{sec:implications} focused on the implications of the signal reported by the EDGES collaboration for IDM models. However, the key analysis explored in this work is applicable to any 21-cm measurement, including the global signal or power spectrum, whether or not it is consistent with EDGES. Thus, it is important to note that several groups have shown that instrumental systematics could give rise to an EDGES-like signal \citep{Hills2018,Singh2019,Bradley2019,Sims2020}, while SARAS 3 recently reported a measurement consistent at $2\sigma$ with the \textit{absence} of an EDGES-like signal in their band \citep{Singh2022}. Additional global signal experiments are coming online now \citep{PRIZM,REACH}, and will soon provide additional insights into the EDGES anomaly.

Beyond the context of EDGES, as shown in Fig.~\ref{fig:model_comp}, DM--baryon interactions have a non-negligible impact on 21-cm signals that may appear to be compatible with $\Lambda$CDM predictions. In particular, the impact of DM--baryon interactions on structure formation can shift absorption signals with $\sim 50-80$ mK depths by more than 20 MHz in frequency. In other words, a firm detection of an absorption trough at low frequencies---regardless of its depth---provides a stringent constraint on Coulomb-like DM--baryon scattering, because it indicates that structure formation could not have been significantly delayed relative to $\Lambda$CDM.

In addition, we note that the degeneracies between astrophysical parameters and IDM models can complicate 21-cm inference. For example, strong DM--baryon interactions that suppress the formation of small halos could be counteracted by efficient star formation in more massive halos, resulting in no net change in the production of photons globally. Furthermore, other DM models beyond the CDM paradigm (e.g., WDM) may delay the onset of the global 21-cm in a similar manner to IDM \cite{Sitwell:2013fpa, Sekiguchi:2014wfa, Lidz:2018fqo, Safarzadeh:2018hhg, Boyarsky:2019fgp, Chatterjee:2019jts, Hibbard2022}, and both may potentially leave subtle imprints on higher-order statistics \cite{Carucci:2015bra}. However, there are reasons to be optimistic about the prospects for overcoming such degeneracies. 

First, while it is easy to imagine that astrophysical processes can ``cancel out'' the effects of IDM at a single redshift, it is much more difficult to explain prolonged effects of Coulomb-like DM--baryon scattering that persist over a wide range of redshifts. Thus, the mass and redshift dependence of structure formation suppression in IDM can only be counteracted with a finely-tuned set of astrophysical parameters. Indeed, the shape of the global 21-cm signal is extremely informative. For example, the delay of structure formation in WDM scenarios can be detected even if there is considerable uncertainty in the values of astrophysical parameters or the astrophysical parameterization itself \citep{Hibbard2022}.

Second, any attempt to counteract suppressed structure with more efficient star formation or photon production (per SFR) in massive halos results in a change in the bias of sources. As a result, 21-cm power spectrum measurements will further help to disentangle DM physics from astrophysics through a combination of redshift and scale-dependent measurements. At a simpler level, power spectrum measurements can test the EDGES scenario directly, given that any excess cooling or excess radio backgrounds will amplify 21-cm fluctuations as well \citep{Fialkov2019}. A number of experiments, e.g., GMRT, LoFAR, MWA, LWA, and HERA are making rapid progress towards 21-cm signal detection that will further test the EDGES anomaly \cite{Paciga2013,Mertens2020,Rahimi2021,Garsden_2021,HERA2022a}.

Finally, as we expand on below, 21-cm observations are not the only probe of signatures of IDM at high redshifts. Measurements of the high-$z$ galaxy population already provide strong constraints on a subset of astrophysical parameters relevant to the 21-cm background \citep{Mirocha2017,Park2019}, and will only improve in the coming years due to surveys with observations from the James Webb Space Telescope (JWST). As a result, it will be increasingly difficult to build \textit{ad hoc} models capable of simultaneously fitting anomalous 21-cm signals and high-$z$ galaxy luminosity functions. Galaxy counts also provide a constraint on the small-scale matter power spectrum \cite{Sabti2022}, and provide yet another window into models that suppress structure formation at high-$z$.

\subsection{Effects Beyond 21-cm Cosmology}\label{sec:other_effects}

\begin{figure}[t!]
    \centering
    \includegraphics[width=\linewidth]{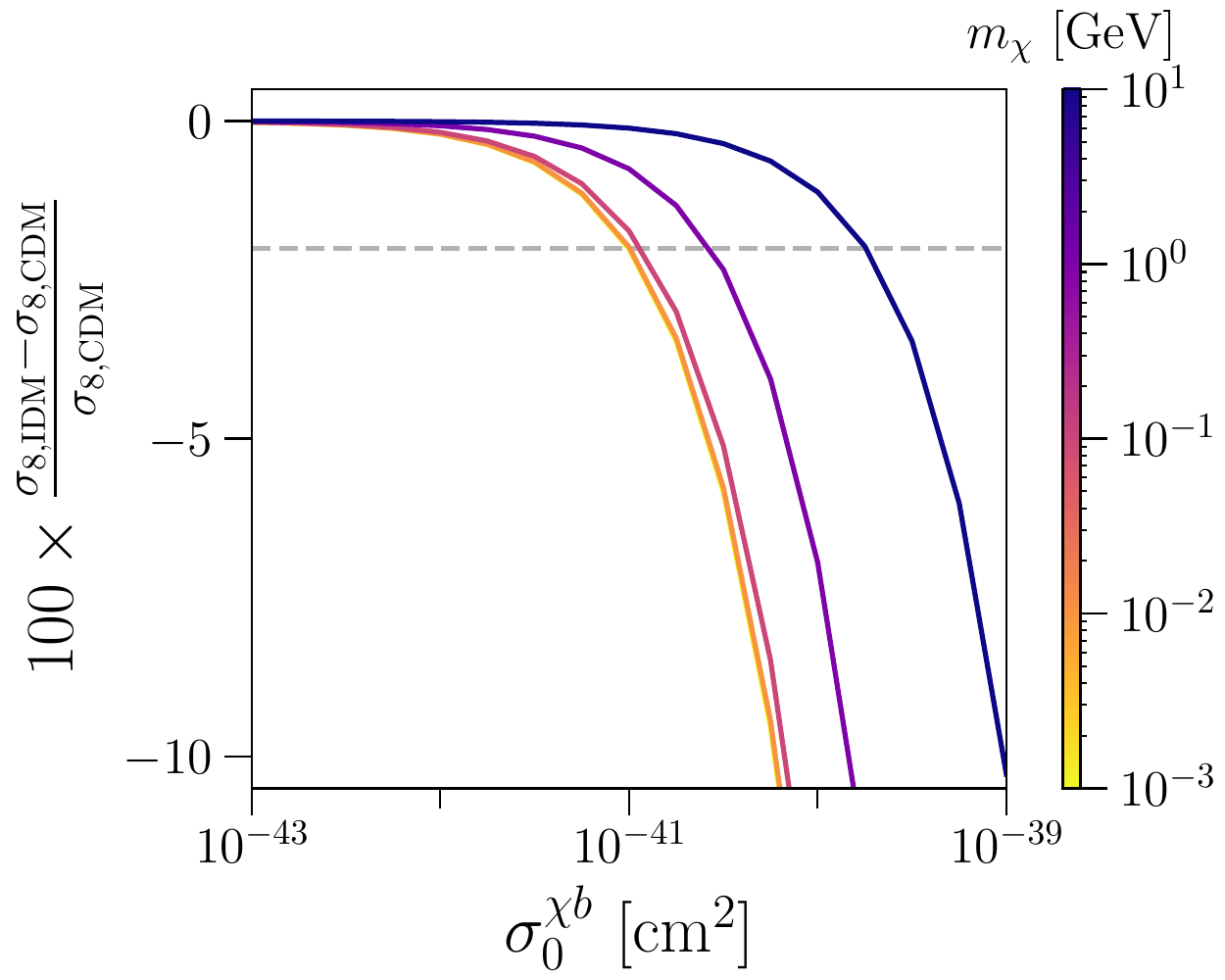}
    \caption{The residual between $\sigma_8$ in Coulomb-like DM--baryon scattering models vs.\ CDM, holding other cosmological parameters fixed. The dashed line approximately corresponds to the $99\%$ confidence-interval lower limit from \textit{Planck} 2018 CDM analysis \cite{Planck2018}. This illustrates the difference between the derived values of $\sigma_{8}$ from assuming a CDM vs. IDM cosmology as a function of the IDM cross section.}
    \label{fig:s8_sigma}
\end{figure}

Beyond the 21-cm signal, the suppression of density perturbations due to momentum exchange in Coulomb-like DM--baryon scattering models is also expected to impact other cosmological observables that are sensitive to the formation and growth of structure. For example, in the linear regime, Fig.~\ref{fig:s8_sigma} summarizes how $\sigma_{8}$---i.e., the mass variance (Eq.~\ref{eq:sigmaM}) smoothed on a scale of $8~\mathrm{Mpc}/h$---responds to the IDM interaction cross section and DM mass, relative to its value in CDM. \textit{Planck} 2018 $99\%$ confidence interval on $\sigma_{8}$ implies an upper bound on the cross section of $\approx 10^{-41}$, for a DM mass of $m_{\chi}=1\,\mathrm{MeV}$, which is comparable to the \textit{Planck} limit on the same model, shown in Fig.~\ref{fig:allowed_gs}. We emphasize that a self-consistent likelihood analysis is needed to infer precise bounds on $\sigma_{8}$ in IDM cosmology. Nonetheless, Fig.~\ref{fig:s8_sigma} clearly illustrates that Coulomb-like DM--baryon interactions alter the linear matter power spectrum at an observable level.

The corresponding suppression of the power spectrum is also interesting in the non-linear regime, given that probes of structure at late times prefer a lower value of $\sigma_8$ than inferred by extrapolating CMB measurements (e.g., see \cite{Abdalla:2022yfr} for a review). However, more detailed modeling of non-linear structure formation in Coulomb-like DM--baryon scattering models, including the potential impact of the interactions on structure at very late times, would be needed to assess the viability of such models to ease the $\sigma_8$ tension. We leave this line of inquiry for future work.

\section{Summary}\label{sec:conclusion}

In this work, we have explored the impact of DM--baryon interactions on structure formation in the context of 21-cm cosmology. In particular, we focused on a popular set of models that include Coulomb-like DM--baryon scattering and millicharged DM,  where momentum-transfer cross section scales with relative particle velocity $v$ as $\sigma(v)=\sigma_{0}v^{-4}$. This model is motivated from the theory perspective, but also represents complex cosmological phenomenology; in particular, this model represents a rich case study for interacting DM, where the interaction-driven heat exchange and the suppression of structure both produce leading-level impact on cosmology. We find that the delay and suppression of structure formation, assumed to be negligible in all previous studies of cosmologies with millicharge, plays a critical role in defining the depth and the timing of the global 21-cm signal from Cosmic Dawn.

We present the first self-consistent modeling of the global signal in cosmologies with $v^{-4}$ interaction, simultaneously evolving the population of DM halos, the background emissions from galaxies at Cosmic Dawn, the ionization history, and the heat exchange between DM and baryons, within interacting DM cosmology. We additionally include improved modeling of the scattering correction to the Wouthuysen-Field coupling strength to the 21-cm spin temperature, accounting for to line profile effects; we find that this correction is essential to correctly predict the global signal in interacting models that cool the gas to temperatures substantially lower than in CDM (i.e., temperatures below $\sim 1\,\mathrm{K}$). 

Our method is schematically represented in Fig.~\ref{fig:pipeline}. We evolved the density perturbations in linear theory using a modified version of \verb|CLASS|. In order to model the non-linear formation of structure, we computed the HMF using a modified version of \verb|Galacticus|. By modifying the 21-cm code \verb|ARES| and including the inputs from \verb|CLASS| and \verb|Galacticus|, we modeled the evolution of the global signal including the effects of both heat and momentum exchange due to DM--baryon interactions. 

We found that, even for relatively moderate interaction strengths allowed by the CMB, the fraction of matter collapsed into halos capable of forming stars is suppressed at the $\mathcal{O}(1)$ level, over the redshift range relevant for the 21-cm signal; for the largest cross sections we consider ($\sigma_0\gtrsim 10^{-40}~\mathrm{cm}^2$), the HMF and corresponding collapsed fraction are suppressed by several orders of magnitude relative to CDM; these results are illustrated in Fig.~\ref{fig:hmf_fcoll}. These effects likely have a measurable impact on a range of structure-formation observables, broadly discussed in Sec.~\ref{sec:discussion}. 

In context of the global 21-cm signal from Cosmic Dawn, our fully self-consistent approach to modeling the effects of DM interactions leads to significant changes in the amplitude and timing of the signal, compared to previous studies; the results are summarize in Fig.~\ref{fig:model_comp}. In Fig.~\ref{fig:allowed_sm}, we illustrate how these corrections to the signal modeling affect the region of the interacting DM parameter space that is consistent with the EDGES measurement from 2018. With a generous consistency criterion, and allowing for wide variations in astrophysical parameters that affect the signal, we find the cross sections necessary to reproduce an EDGES-like amplitude would simultaneously delay the signal far later than the detected signal. Therefore, we conclude that 100\% millicharge models cannot reproduce the EDGES measurement in any part of DM parameter space. This result stands in stark contrast to previous studies that neglect the effects of DM interactions on structure formation and the effects of Lyman-$\alpha$ coupling corrections for a cold IGM; it is also entirely independent of other observational bounds.

Finally, in Sec.~\ref{sec:discussion}, we discuss the implications of our findings in the context of future 21-cm measurements and other cosmological probes of DM. We conclude that the effects of heat exchange between DM and baryons cannot be considered in isolation from the effects of these interactions on structure formation, even in models that mainly feature late-time scattering. Our analysis critically informs future studies of DM microphysics with cosmological observables, and informs the interpretation of data from future 21-cm experiments and galaxy surveys.

\section*{Acknowledgements}
VG acknowledges the support from the National Science Foundation (NSF) under Grant No. PHY-2013951 and from NASA through the Astrophysics Theory Program, Award Number 21-ATP21-0135. 
KB acknowledges support from the NSF under Grant No.~PHY-2112884.

\bibliography{main}

\end{document}